\documentclass{aa} % for a referee version
\usepackage{graphicx}
\usepackage{txfonts}
\newcommand{\DSPSR}{\texttt{DSPSR}}
\newcommand{\PRESTO}{\texttt{PRESTO}}

\newcommand{\SIGPROC}{\texttt{SIGPROC}}
\newcommand{\PSRCHIVE}{\texttt{PSRCHIVE}}
\newcommand{\TEMPO}{\texttt{TEMPO}}

\newcommand{\PULSARMINER}{\texttt{PULSAR\_MINER}}

\newcommand{\dmunit}{pc\,cm$^{-3}$}

\newcommand{\rmunit}{{\rm rad~m}$^{-2}$}
\newcommand{\msun}{M$_{\sun}$}

\newcommand{\Pobs}{P_{\rm obs}}

\newcommand{\Pb}{P_{\rm b}}

\newcommand{\fc}{f_{\rm c}}

\newcommand{\Mp}{M_{\rm p}}
\newcommand{\Mc}{M_{\rm c}}

\newcommand{\xp}{x_{\rm p}}

\newcommand{\rcore}{r_{\rm c}}

\newcommand{\rhalfmass}{r_{\rm hm}}

\newcommand{\h}{^{\rm h}}
\newcommand{\m}{^{\rm m}}
\newcommand{\s}{^{\rm s}}

\begin{document}

   \title{TRAPUM discovery of thirteen new pulsars in NGC~1851 using MeerKAT}

   \subtitle{}
   \titlerunning{Thirteen new pulsars in NGC~1851}

   \author{
A.~Ridolfi\inst{\ref{1},\ref{2}}\thanks{E-mail: alessandro.ridolfi@inaf.it}
\and P.~C.~C.~Freire\inst{\ref{2}}\thanks{E-mail: pfreire@mpifr-bonn.mpg.de}
\and T.~Gautam\inst{\ref{2}}
\and S.~M.~Ransom\inst{\ref{3}}
\and E.~D.~Barr\inst{\ref{2}}
\and S.~Buchner\inst{\ref{4}}
\and M.~Burgay\inst{\ref{1}}
\and F.~Abbate\inst{\ref{2}}
\and V.~Venkatraman~Krishnan\inst{\ref{2}}
\and L.~Vleeschower\inst{\ref{5}}
\and A.~Possenti\inst{\ref{1},\ref{6}}
\and B.~W.~Stappers\inst{\ref{5}}
\and M.~Kramer\inst{\ref{2},\ref{5}}
\and W.~Chen\inst{\ref{2}}
\and P.~V.~Padmanabh\inst{\ref{2},\ref{7}}
\and D.~J.~Champion\inst{\ref{2}}
\and M.~Bailes\inst{\ref{8},\ref{9}}
\and L.~Levin\inst{\ref{5}}
\and E.~F.~Keane\inst{\ref{10}}
\and R.~P.~Breton\inst{\ref{5}}
\and M.~Bezuidenhout\inst{\ref{5}}
\and J.-M.~Grie{\ss}meier\inst{\ref{11},\ref{12}}
\and L.~K\"{u}nkel\inst{\ref{13}}
\and Y.~Men\inst{\ref{2}}
\and F.~Camilo\inst{\ref{4}}
\and M.~Geyer\inst{\ref{4}}
\and B.~V.~Hugo\inst{\ref{4},\ref{14}}
\and A.~Jameson\inst{\ref{8},\ref{9}}
\and A. Parthasarathy\inst{\ref{2}}
\and M.~Serylak\inst{\ref{15},\ref{16}}
          }

   \institute{
INAF -- Osservatorio Astronomico di Cagliari, Via della Scienza 5, I-09047 Selargius (CA), Italy\label{1}
\and
Max-Planck-Institut f\"{u}r Radioastronomie, Auf dem H\"{u}gel 69, D-53121 Bonn, Germany\label{2}
\and
National Radio Astronomy Observatory, 520 Edgemont Rd., Charlottesville, VA 22903, USA\label{3}
\and
South African Radio Astronomy Observatory (SARAO), 2 Fir Street, Black River Park, Observatory, Cape Town, 7925, South Africa\label{4}
\and
Jodrell Bank Centre for Astrophysics, Department of Physics and Astronomy, The University of Manchester, Manchester M13 9PL, UK\label{5}
\and
Universit\`a di Cagliari, Dipartimento di Fisica,  S.P. Monserrato-Sestu Km 0,700,  I-09042 Monserrato (CA), Italy \label{6}
\and
Max-Planck-Institut f\"{u}r Gravitationsphysik (Albert-Einstein-Institut), 30167 Hannover, Germany \label{7}
\and
Centre for Astrophysics and Supercomputing, Swinburne University of Technology, P.O. Box 218, Hawthorn, VIC 3122, Australia\label{8}
\and
ARC Centre of Excellence for Gravitational Wave Discovery (OzGrav)\label{9}
\and
School of Physics, Trinity College Dublin, College Green, Dublin 2, Ireland\label{10}
\and
LPC2E - Universit\'{e} d'Orl\'{e}ans /  CNRS, 45071 Orl\'{e}ans cedex 2, France\label{11}
\and
Station de Radioastronomie de Nan\c{c}ay, Observatoire de Paris, PSL Research University, CNRS, Univ. Orl\'{e}ans, OSUC, 18330 Nan\c{c}ay, France\label{12}
\and
Fakult\"{a}t f\"{u}r Physik, Universit\"{a}t Bielefeld, Postfach 100131, D-33501 Bielefeld, Germany\label{13}
\and
Department of Physics and Electronics, Rhodes University, PO Box 94, Grahamstown 6140, South Africa\label{14}
\and
SKA Observatory, Jodrell Bank, Lower Withington, Macclesfield, SK11 9FT, United Kingdom\label{15}   \and
Department of Physics and Astronomy, University of the Western Cape, Bellville, Cape Town, 7535, South Africa\label{16}
                }

   \date{}

% \abstract{}{}{}{}{}
% 5 {} token are mandatory
 
  \abstract{ We report the discovery of 13 new pulsars in the globular cluster NGC 1851 by the TRAPUM Large Survey Project using the MeerKAT radio telescope. The discoveries consist of six isolated millisecond pulsars (MSPs) and seven binary pulsars, of which six are MSPs and one is mildly recycled. For all the pulsars, we present the basic kinematic, astrometric, and orbital parameters, where applicable, as well as their polarimetric properties, when these are measurable. Two of the binary MSPs (PSR~J0514$-$4002D and PSR~J0514$-$4002E) are in wide and extremely eccentric ($e>0.7$) orbits with a heavy white dwarf and a neutron star as their companion, respectively.
With these discoveries, NGC~1851 is now tied with M28 as the cluster with the third largest number of known pulsars (14). 
Its pulsar population shows remarkable similarities with that of M28, Terzan 5 and other clusters with comparable structural parameters. The newly-found pulsars are all located in the innermost regions of NGC~1851 and will likely enable, among other things, detailed studies of the cluster structure and dynamics.}


   \maketitle
%
%________________________________________________________________

\begin{table*}
\renewcommand{\arraystretch}{1.1}
\setlength{\tabcolsep}{0.15cm}
\footnotesize
\centering
\caption{List of the MeerKAT observations of NGC~1851 recorded for this work. All PTUSE data were acquired with real-time coherent de-dispersion, using a DM of 52.1\,\dmunit. $t_{\rm samp}$: sampling time; $N_{\rm pol}$: number of Stokes parameters; $f_{\rm c}$: central frequency; $\Delta f$: observing bandwidth; $N_{\rm chan}$: number of frequency channels; $N_{\rm ant}$: number of antennas; $N_{\rm beam}$: number of tied-array beams. Before being searched for new pulsars, all the APSUSE observations were incoherently dedispersed at a DM of 52.14\,\dmunit\ and subbanded, resulting in 256-channel filterbank files, while all the PTUSE observations were summed in polarization and decimated in both time and frequency, retaining total-intensity only, 512-channels and $75.29~\mu$s time resolution (see Section \ref{sec:data_reduction}). The native-resolution full-Stokes PTUSE data were used for polarimetric measurements (see Section \ref{sec:polarimetry}).}
\label{tab:list_observations}
\begin{tabular}{crrrrrcccccr}
\hline
Obs. id            & \multicolumn{1}{c}{Start Time}    & \multicolumn{1}{c}{Start Time}       & \multicolumn{1}{c}{Length}    & \multicolumn{1}{c}{Backend} & \multicolumn{1}{c}{$t_{\rm samp}$} & \multicolumn{1}{c}{$N_{\rm pol}$}& \multicolumn{1}{c}{$f_{\rm c}$}  & \multicolumn{1}{c}{$\Delta f$}  & \multicolumn{1}{c}{$N_{\rm chan}\phantom{*}$} & \multicolumn{1}{c}{$N_{\rm ant}$}  & $N_{\rm beam}$    \\
               & \multicolumn{1}{c}{(Date)}        & \multicolumn{1}{c}{(MJD)}            & \multicolumn{1}{c}{(s)}  & \multicolumn{1}{c}{}  & \multicolumn{1}{c}{($\mu$s)} &  & (MHz)        & (MHz) &               &   \\
\hline
01L     & 15 Jan 2021 & 59229.626 & 14400 & APSUSE & 76.56 & 1 & 1284 & 856 &  4096$^*$ & 64   & 288 \\
02L     & 08 May 2021 & 59342.278 &  7200 & APSUSE & 76.56 & 1 & 1284 & 856 &  4096$^*$ & 60   & 288 \\
03L     & 08 May 2021 & 59342.586 &  7200 & APSUSE & 76.56 & 1 & 1284 & 856 &  4096$^*$ & 60   & 288 \\
04U     & 21 May 2021 & 59355.253 &  7200 & APSUSE & 60.24 & 1 &  816 & 544 &  4096$^*$ & 56   & 277 \\
05U     & 21 May 2021 & 59355.628 &  7200 & APSUSE & 60.24 & 1 &  816 & 544 &  4096$^*$ & 60   & 277 \\
06U     & 24 May 2021 & 59358.242 &  7200 & APSUSE & 60.24 & 1 &  816 & 544 &  4096$^*$ & 56   & 277 \\
07U     & 26 May 2021 & 59360.607 &  7200 & APSUSE & 60.24 & 1 &  816 & 544 &  4096$^*$ & 56   & 277 \\
08U     & 03 Jul 2021 & 59398.485 &  7200 &  PTUSE &  9.41 & 4 &  816 & 544 & 1024$\phantom{*}$ & 61 &   1 \\
09U     & 17 Aug 2021 & 59443.233 &  7200 &  PTUSE &  9.41 & 4 &  816 & 544 & 1024$\phantom{*}$ & 59 &   1 \\
10U     & 01 Sep 2021 & 59458.067 &  7200 &  PTUSE &  9.41 & 4 &  816 & 544 & 1024$\phantom{*}$ & 57 &   1 \\
11U     & 01 Oct 2021 & 59487.942 &  7200 &  PTUSE &  9.41 & 4 &  816 & 544 & 1024$\phantom{*}$ & 61 &   1 \\
12U     & 07 Nov 2021 & 59525.075 &  10800 &  PTUSE &  9.41 & 4 &  816 & 544 & 1024$\phantom{*}$ & 61 &  1 \\
13U     & 09 Nov 2021 & 59527.919 &  7200 &  PTUSE &  9.41 & 4 &  816 & 544 & 1024$\phantom{*}$ & 56 &   1 \\
\hline
\end{tabular}\\
\end{table*}

\section{Introduction}
\label{sec:intro}

Globular clusters (GCs) have historically been popular targets for radio pulsar surveys \citep[e.g.][]{Ransom2008}. This is due to their nature, which makes them extremely fecund pulsar seedbeds. GCs are ancient (ages of several Gyr) stellar systems that are kept together by their own gravity, resulting in a spherical distribution of stars. In their cores the stellar number density can reach up to $10^{3-6}$\,pc$^{-3}$, i.e., up to $\sim 10^6$ times the typical values found in the Solar neighbourhood.

In such crowded environments it is fairly common that an old, ``dead'' neutron star (NS) gravitationally interacts with binary systems within the cluster. If the encounter is close enough, this can cause the ejection of the lighter member of the previous binary and the formation of a new binary system, consisting of a NS and a main sequence (MS) star \citep{Hills1975}. The MS star then evolves, fills its Roche lobe and starts transferring matter to the NS, which can thus be ``recycled''. During this phase the system is seen as a low-mass X-ray binary (LMXB). Many of these systems then evolve to form millisecond pulsars (MSPs, e.g. \citealt[][]{Alpar+1982,Radhakrishnan_Srinivasan1982,Papitto+2013}). The formation of LMXBs by dynamical means is the reason why there are, by unit of stellar mass, about three orders of magnitude more LMXBs and MSPs in GCs than in the Galactic disk \citep{Clark75,vandenBerg2020}.

Searches for pulsars in GCs are first and foremost limited by sensitivity: owing to the large distances (in most cases $>5$\,kpc) of these stellar systems, only the brightest pulsars can be detected. After a decade characterized by a slow discovery rate, mainly due to a sensitivity limit being reached at the existing radio telescopes, a new surge in the total number of pulsars in GCs has recently occurred. In the last three years, the total GC pulsar population has increased by more than 50 per cent, reaching, at the time of writing, 235 known pulsars\footnote{See \url{https://www3.mpifr-bonn.mpg.de/staff/pfreire/GCpsr.html} for the most up-to-date number.} in 36 different clusters. While a few of the new discoveries resulted from instrumentation upgrades made at the Parkes \citep{Dai+2020}, and GMRT radio telescopes (Gautam et al., submitted), the vast majority were found by the two newly-built major radio facilities: the Chinese Five-hundred-meter Aperture Spherical Telescope (FAST, \citealt{Nan+2011}) and the South African MeerKAT radio telescope \citep{Jonas+2016,Camilo2018}. Despite their very different designs and total collecting areas, both FAST and MeerKAT are providing a major leap in the available raw sensitivity for pulsar searching experiments in their respective hemispheres.

As a result, since its first GC observation taken in 2018, FAST has discovered over thirty new pulsars in GCs \citep{Wang+2020,Pan+2020,Pan+2021a,Pan+2021b,Qian_Pan2021,Yan+2021}, some of which are in clusters with no previously known pulsars.

Similarly, MeerKAT began observing GCs in 2019, with a first ``census'' campaign carried out in the context of the MeerTime\footnote{\url{http://www.meertime.org}} \citep{Bailes+2020} Large Survey Project (LSP). During this, a sample of nine promising GCs hosting previously known pulsars were observed. These observations used the Pulsar Timing User Supplied Equipment (PTUSE) backend, which, at the time, could record a single tied array beam on the sky. Hence those observations used only the central 44 antennas of the array, to trade off some sensitivity for an increased field of view. These data served a variety of purposes, among which were to: a) test the telescope performance; b) time and characterize previously known pulsars \citep{Abbate+2020a}; c) perform a first search for new pulsars. The latter turned out to be successful, with eight new MSPs found across six different clusters \citep{Ridolfi+2021}.

Besides being scientifically very fruitful, the MeerKAT GC census was also key in planning future observations and paved the way for a more thorough GC pulsar survey, which started in mid 2020 as part of the TRansients And PUlsars with MeerKAT (TRAPUM\footnote{\url{http://www.trapum.org}}) LSP \citep{Stappers_Kramer2016}. Unlike the census observations, the TRAPUM GC pulsar survey uses the Filterbanking Beamformer User Supplied Equipment (FBFUSE) and the Accelerated Pulsar Search User Supplied Equipment (APSUSE) computing clusters as backend to form up to 288 tied-array beams on the sky, tiled radially out from the GC centre using an optimal hexagonal tessellation. This means that we can use up to all of the 64 MeerKAT antennas, substantially increasing the sensitivity compared to the initial GC census. This survey also extends the sample of target GCs to a total of 28, and includes observations using different frequency bands, depending on the distance to the cluster. One of these is the relatively unexplored Ultra High Frequency (UHF) band. The latter, which at MeerKAT operates between 544 and 1088 MHz, is proving to be particularly useful, as shown in this work.

One of the GCs selected for the TRAPUM GC survey is NGC~1851, which is the main focus of this paper. This is a fairly compact cluster, having a core radius of $\rcore=0.09$\,arcmin and a half-mass radius of $\rhalfmass=0.85$\,arcmin \citep{Miocchi+2013}. With centre coordinates being at right ascension $\alpha_{\rm J2000}=$ 05$\h$\,14$\m$\,06.76$\s$ and declination $\delta_{\rm J2000}=-40\degr\,02\arcmin\,47.6\arcsec$ \citep{Harris2010}, it is located in the Southern constellation of Columba at a distance of $11.95\pm0.13$~kpc \citep{Baumgardt_Vasiliev2021}, far below the Galactic plane (at Galactic coordinates $l=244^\circ.5$, $b=-35^\circ.0$). Until recently, NGC~1851 was known to be the host of a single, but very peculiar binary pulsar, PSR~J0514$-$4002A (also known as NGC~1851A). Discovered by \citet{Freire+2004} at 350~MHz with the GMRT, and subsequently timed with the Green Bank Telescope (GBT) and the GMRT at similar frequencies \citep{Freire+2007,Ridolfi+2019}, NGC~1851A is a 4.99-ms pulsar in a wide and extremely eccentric ($e=0.89$) orbit. This characteristic made it possible to precisely measure two post-Keplerian parameters, which in turn led to a precise determination of the system component masses, something that is rarely possible. NGC~1851A is most likely the result of a secondary exchange interaction, where the original low-mass star that recycled the pulsar was replaced by the current, massive ($\Mc=1.22$\,\msun) companion. The existence of such a peculiar binary motivated our searches in this cluster, as it informed our suspicions that other such systems may be present therein.

Here we present the results of our search for new pulsars in NGC~1851 using MeerKAT observations made in the context of the TRAPUM GC pulsar survey, as well as follow-up observations taken within the MeerTime programme. In Section \ref{sec:observations} we describe the observations, their instrumental set-ups and the data analysis methods. We present the results in Section \ref{sec:results} and discuss their implications in Section \ref{sec:discussion}. Finally, in Section \ref{sec:conclusions}, we summarize our findings and examine possible scientific developments.

%%%%%%%%%%%%%%%%%%%%%%%%%%%%%%%%%%%%%%%%%%%%%%%%%%%%%%%
%   OBSERVATIONS
%%%%%%%%%%%%%%%%%%%%%%%%%%%%%%%%%%%%%%%%%%%%%%%%%%%%%%%

\begin{figure*}
\centering
	\includegraphics[width=0.47\textwidth]{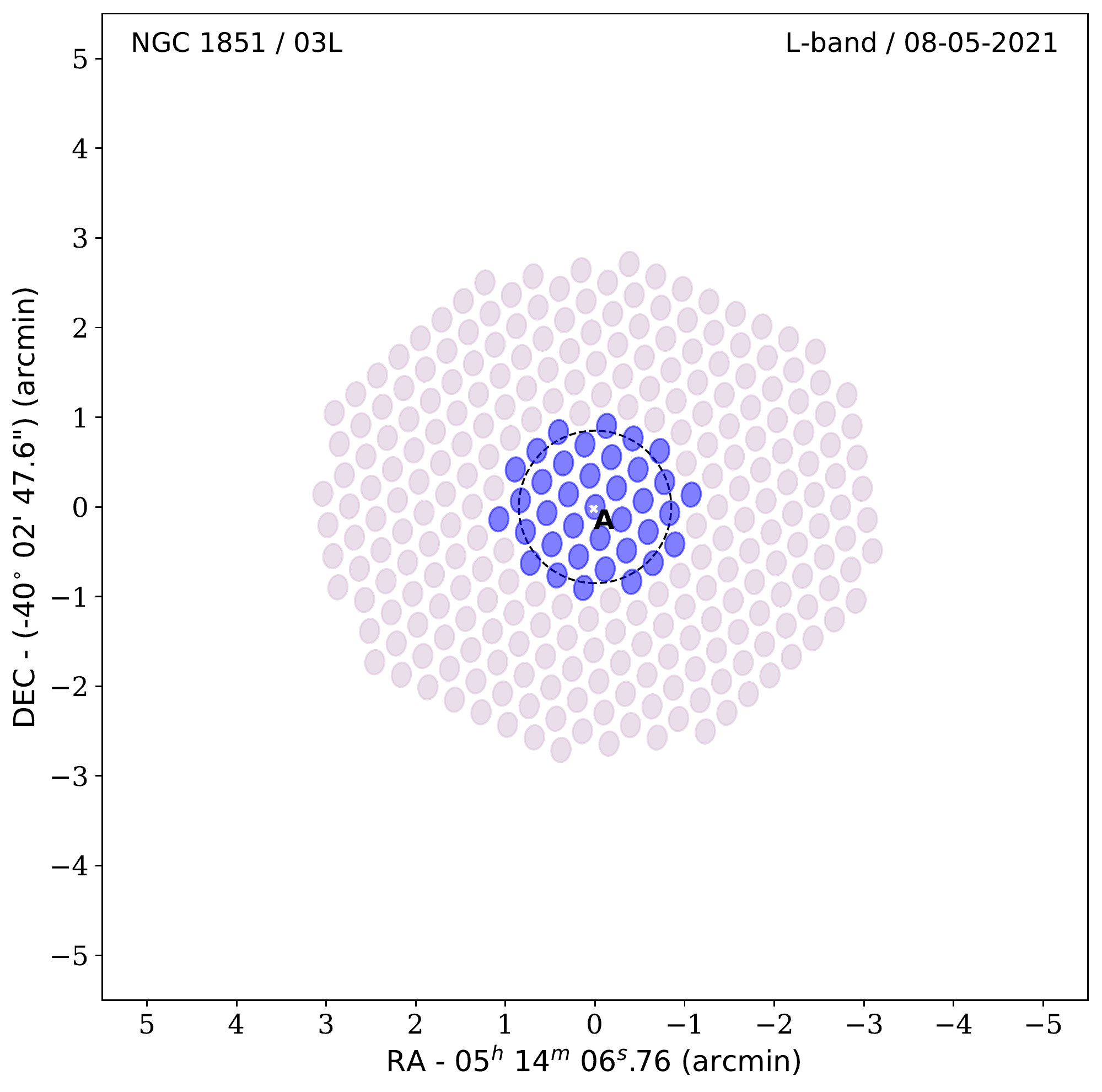}
	\qquad
	\includegraphics[width=0.47\textwidth]{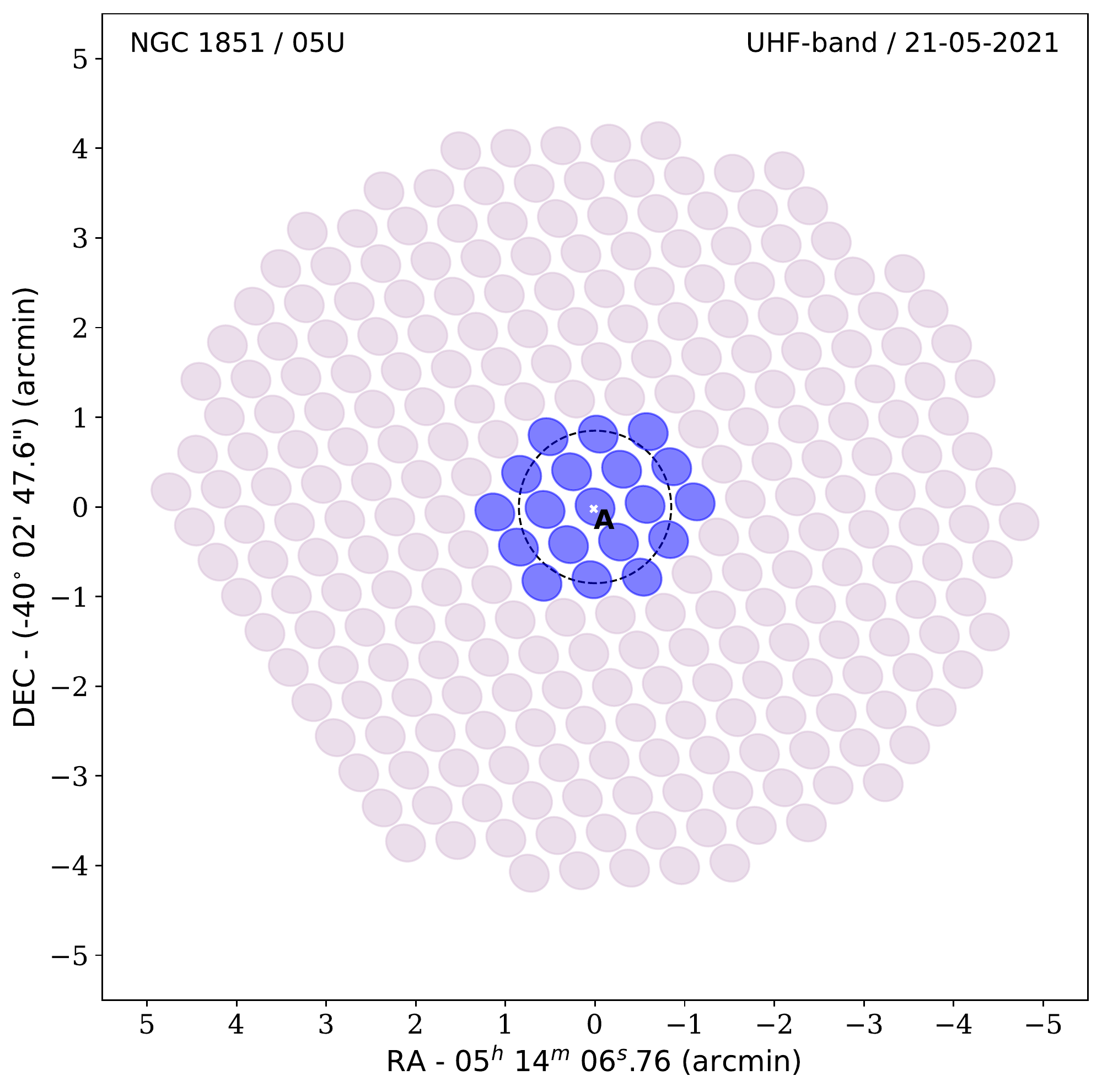}
	\caption{TRAPUM tilings of two observations of NGC~1851, performed with MeerKAT on 2021 May 8 at L-band (left) and on 2021 May 21 at UHF-band (right), with similar telescope elevations.  Both panels show the same portion of the sky, for an easier comparison of the area covered by the tilings in the two different bands. The ellipses show the tied-array beams up to 84 per cent of their boresight power at their respective central frequencies, i.e. 816 MHz for the UHF band, 1284 MHz for the L band. The blue ellipses show the beams that have been searched for this work. The dashed circle shows the half-mass radius of NGC~1851, located at 0.85 arcmin from the cluster centre. The position of the previously known pulsar, NGC~1851A, is shown by the white cross.}
  	\label{fig:tiling_patterns}
\end{figure*}

\section{Observations and data analysis}
\label{sec:observations}

We have observed NGC~1851 with MeerKAT on 13 occasions, from January to November 2021 (see Table \ref{tab:list_observations} for the full list of observations), each time using at least 56 antennas.
The first seven observations were carried out as part of the TRAPUM GC pulsar survey, hence, we recorded the total intensity only with the APSUSE computing cluster as the main backend. The first three of them were made using the L-band receivers, at a central frequency $\fc = 1284$\,MHz and with a bandwidth $\Delta f=856$\,MHz. The latter was split into 4096 channels and sampled every $76.56\,\mu$s.  The other four TRAPUM observations were made with the UHF receivers, at a central frequency $\fc = 816$\,MHz and with $\Delta f =544$\,MHz, also split into 4096 channels but sampled every $60.24\,\mu$s. In all the TRAPUM observations, the computing power of FBFUSE was used to synthesize either 277 or 288 tied-array beams \citep{Chen+2021}. These allowed us to cover an area of several square arcminutes around the nominal centre of NGC~1851, which extended a few times beyond the half-mass radius of the cluster (see Figure \ref{fig:tiling_patterns}).

The other six observations were carried out as part of the MeerTime GC pulsar timing programme.  Their set-up was chosen based on the results obtained from the TRAPUM data previously taken: we used the UHF receivers and PTUSE as the backend. The latter produced a single tied-array beam, synthesized at the centre of NGC~1851. This was enough to observe all the new pulsars at once with maximum sensitivity (see Section \ref{sec:results}). The observing band was divided into 1024 channels, coherently de-dispersed at a dispersion measure (DM) of 52.1\,\dmunit\ (i.e. the DM of the only pulsar previously known in the cluster), and recorded in full-Stokes mode, every $9.41\,\mu$s. Since the PTUSE data were also stored as search-mode files, they were used to confirm and follow-up the pulsars found in the TRAPUM data, and also to search for new ones.

For some of our analyses, we also took advantage of the large set of archival observations of NGC~1851 taken with the GBT and the GMRT telescopes. The GBT data were taken with central frequencies of 350, 820 and 1950 MHz in the years 2004--2006 and are described by \citet{Freire+2007}. The GMRT data were acquired with a central frequency of 400 MHz from 2017 to 2020, with observational set-ups that are described by \citet{Ridolfi+2019}. 

\subsection{Sensitivity}
Given the observing parameters described above, we can use the radiometer equation \citep{Dewey+1985} to estimate our typical sensitivity: we find that we are sensitive to pulsars with minimum mean flux densities of $S_{\rm L}=10~\mu$Jy and $S_{\rm UHF}=14~\mu$Jy in the L band and UHF band, respectively. These are estimated considering, for both observing bands: 60 antennas, for a total telescope gain of 2.63~K\,Jy$^{-1}$; 2 hours of integration time; two polarizations summed; a correction factor of 1.1 to account for possible digitization and other losses of various nature; a pulse duty cycle of 8 per cent; a DM of 52~\dmunit; a detection signal-to-noise ratio (S/N) of 10. For the L-band observations, we assume an effective bandwidth (i.e. usable bandwidth after the removal of radio frequency interference, henceforth RFI) of 650 MHz, and a total system temperature of 26~K\footnote{\url{https://skaafrica.atlassian.net/rest/servicedesk/knowledgebase/latest/articles/view/277315585}}. The latter is the sum of the receiver temperature ($T_{\rm rec}=18$~K), and contributions from the sky ($T_{\rm sky}=3.5$~K), atmosphere ($T_{\rm atm}=2$~K) and ground spillover ($T_{\rm spill}=2.5$~K). For the UHF observations, we consider an effective bandwidth of 500~MHz, and a total system temperature of $31.5$~K (with the contributions being $T_{\rm rec}=20$~K, $T_{\rm sky}=5.5$~K, $T_{\rm atm}=1.5$~K, $T_{\rm spill}=4.5$~K). For extremely fast-spinning pulsars, with spin periods $\lesssim 2$~ms, the minimum detectable mean flux densities are higher because of pulse smearing effects, in turn due to the sampling time and finite channel width becoming important.

\subsection{Data reduction}
\label{sec:data_reduction}
Before being analyzed, all the observations were pre-processed. 
In particular, all the TRAPUM beams, which are recorded as \SIGPROC's\footnote{\url{http://sigproc.sourceforge.net}} ``filterbank'' search-mode files, were first incoherently de-dispersed at the nominal cluster DM of 52.1\,\dmunit. Then, their frequency channels were summed in groups of 16, bringing their total number from 4096 to 256, substantially reducing the total data volume. While doing so, the Inter-Quartile Range Mitigation algorithm\footnote{\url{https://github.com/v-morello/iqrm}} \citep{Morello+2021} was applied to significantly reduce the presence of RFI in the data. This algorithm is very effective with the MeerKAT data in both the L and UHF bands. The resulting files were, in the vast majority of cases, almost completely free of RFI, greatly easing their subsequent analysis. Once the processed files were produced and checked, the original 4096-channel data were deleted.
The MeerTime PTUSE observations, which are recorded as PSRFITS \citep{Hotan+2004} search-mode files, were decimated by a factor of two in the frequency domain (preemptively applying appropriate shifts to the channels, to account for the cluster DM of 52.1\,\dmunit), a factor of eight in the time domain, and the two polarizations were summed. This was done with \texttt{psrfits\_subband}, one of the tools of the \texttt{PSRFITS\_UTILS}\footnote{\url{https://github.com/scottransom/psrfits\_utils}} package. The resulting files, which therefore have 512 channels, a time resolution of $75.29\,\mu$s and total-intensity only information, were used to search for new pulsars.
The original full-Stokes, native-resolution PTUSE data were instead used to carry out polarimetric measurements, and they will also be preserved for future high time resolution timing and more in-depth polarization studies.

\subsection{Search}
All the thirteen NGC~1851 observations were searched for new pulsars using \PULSARMINER\footnote{\url{https://github.com/alex88ridolfi/PULSAR\_MINER}}, an automated pipeline based on the \PRESTO\footnote{\url{https://github.com/scottransom/presto}} pulsar searching package \citep{Ransom+2002,Ransom2011}. Each search-mode file was cleaned using  \PRESTO' \texttt{rfifind} and then searched in the Fourier domain with \PRESTO's \texttt{accelsearch}. The latter can recover a pulsar's periodic signal in a situation where its Doppler shift is varying rapidly because of its orbital motion\footnote{We recall that the signal of a pulsar with a spin period $P$, orbiting in a binary system, will experience a drift in its observed spin frequency by $z = \Delta t_{\rm obs}^2 a_l / (cP)$ bins in the Fourier domain, where $\Delta t_{\rm obs}$ is the length of the observation, $a_l $ is the pulsar's line-of sight component of the orbital acceleration at the time of the observation and $c$ is the speed of light \citep{Ransom+2002}.}, by considering signals whose frequency changes by up to a number $z_{\rm max}$ of Fourier bins around a given frequency. Therefore, we ran \texttt{accelsearch} with a $z_{\rm max}=0$ to target isolated pulsars, and  with a $z_{\rm max}=200$ to target binary pulsars, in both cases summing up to 8 harmonics. This was done on the full-length data, as well as on segments of 120 (for the 4-hour observation), 60 and 30 minutes, with DM trialsspanning the range 49.5$-$54.5\,\dmunit. We refer to Section 3 of \citet{Ridolfi+2021} for further details on all the steps of the searching pipeline. Due to the huge data volume and consequent computational time required, in the case of the TRAPUM observations, we restricted our initial search to only those beams that covered the half-mass radius of NGC~1851 (Figure \ref{fig:tiling_patterns}).
If the same candidate was found in more than one epoch and in beams with compatible sky positions, it was marked as a ``confirmed'' pulsar and further inspected. 

Finally, in addition to the periodicity search, we carried out a single-pulse search. This was done by running \PRESTO's \texttt{single\_pulse\_search.py} with default parameters on all the full-length de-dispersed time series produced by \PULSARMINER\ and by visually inspecting the resulting diagnostic plots.

\subsection{Binary orbits and timing}
\label{sec:binary_orbits}
A high fraction of the confirmed pulsars exhibited some non-zero acceleration, a likely sign of being part of a binary system. For these pulsars, we first tried to determine their rough orbital parameters using the following methods. The first attempt was made using the Period-Acceleration diagram technique described in \citet{Freire+2001}. After that, we fitted the observed spin period as a function of time, $\Pobs(t)$, to an orbital model, with initial parameters derived from the previous method. The $\Pobs(t)$ fitting greatly benefitted from the six closely-spaced observations taken in May 2021 and was implemented using \PRESTO's \texttt{fit\_circular\_orbit.py} for the circular systems, and either \PRESTO's \texttt{fitorb.py} or the FORTRAN code \texttt{fitorbit}\footnote{\url{https://github.com/vivekvenkris/fitorbit}} for the eccentric systems. The  derived orbital parameters were later further refined through timing (see below).

To better characterize all the newly found pulsars, we extracted times-of-arrival (ToAs) from the available detections and did a timing analysis. We did so by first folding the search-mode data with the \DSPSR\footnote{\url{http://dspsr.sourceforge.net}} pulsar software package \citep{vanStraten_Bailes2011}. For the isolated pulsars, we constructed an ephemeris that contained the best barycentric spin period, $P$, and DM as determined by the search pipeline. For the binary pulsars, the ephemeris also included the best-fitting orbital model as previously derived. Using this to fold all the observations often resulted in additional faint detections of the pulsar that were not found by the search pipeline.

The resulting folded archives were processed with \PSRCHIVE\footnote{\url{http://psrchive.sourceforge.net}} \citep{Hotan+2004,vanStraten+2012} to extract topocentric ToAs from each detection. This was done by cross-correlating the observed pulse profiles with an analytical pulse template of the pulsar. These ToAs were then fitted to an appropriate timing model with the \TEMPO\footnote{\url{http://tempo.sourceforge.net}} pulsar timing software. Because of the faintness and the sparsity of the detections of several of the new pulsars, we used \texttt{DRACULA}\footnote{\url{https://github.com/pfreire163/Dracula}} \citep{Freire_Ridolfi2018} to determine the exact number of rotations of the pulsar between groups of ToAs. The use of this algorithm showed that, for most pulsars, many possible solutions can still be obtained using the existing ToA sets. This means that additional observations are required for the determination of the correct timing solutions. For this reason, the timing analysis of these pulsars will be presented elsewhere.

\subsection{DM and flux densities}
\label{sec:flux_densities}
For each pulsar, we ran \PSRCHIVE's \texttt{pdmp} on all the available UHF detections to obtain, among other things, a best-fitting DM value for each of them. The average of the best-fitting DMs was calculated to give the best estimate of the DM associated with the pulsar. The standard error on the average of the same measurements was adopted as the associated uncertainty.

As described in Section \ref{sec:results_dms_fluxes}, the pulsars in NGC~1851 show significant variations in their apparent brightness from epoch to epoch and even within the same observation. Therefore, a single measurement of the mean flux density $S$ for a pulsar, is not representative of its true intrinsic value. Rather, to obtain a more realistic estimate of $S$, we measured the latter from each observation and considered the average value $\langle S \rangle$, and its standard deviation as associated uncertainty. This was done using only the UHF-band observations, since they are the most numerous ones and have the highest detection rate for all the discoveries. For each observation we took the S/N value estimated by \texttt{pdmp}, and used the radiometer equation to calculate the corresponding mean flux density in the UHF band, $S_{\rm UHF}$, for all the detections. The non-detections were also taken into account by arbitrarily assigning $S_{\rm UHF}=7~\mu$Jy to them, as this is exactly half of the estimated minimum detectable $S$ for the UHF band.

\subsection{Polarimetric properties}
\label{sec:polarimetry}
We used the six most recent UHF observations listed in Table \ref{tab:list_observations}, which were taken with PTUSE in full-Stokes mode, to measure the polarimetric properties of the pulsars in NGC~1851. After folding the native-resolution data for each pulsar with \DSPSR, we used \PSRCHIVE's \texttt{pac} routine on the folded archives to correct for the time-varying parallactic angle. This was sufficient to obtain fully calibrated folded archives for the latest four observations (obs. ids 10U, 11U, 12U, 13U), as these were calibrated, correcting for the instrumental response, directly while being recorder. The other two observations (obs. ids 08U and 09U) were not calibrated during the recording, hence we applied a calibration solution accounting for the system's response \emph{a posteriori} on the folded archives \citep{Serylak+2021}. For each pulsar, we then selected the best detections of the sample and summed them together with \PSRCHIVE's \texttt{psradd}, in order to maximize the S/N of the resulting archive. We then ran \PSRCHIVE's \texttt{rmfit} on the latter to measure the associated rotation measure (RM): the code brute-force searched for the best RM within a specified range (which we chose to be between $-500$ and $500$ \rmunit) that maximizes the measured fraction of linearly-polarized flux density.  If the pulsar was sufficiently polarized and the S/N high enough, \texttt{rmfit} returned the best RM and its uncertainty, evaluated by fitting a Gaussian function to the polarized flux as a function of trial RM. The so obtained RM value was used to de-Faraday the high-S/N summed archive of the pulsar, and to plot the resulting linearly- an circularly-polarized integrated profiles.

\subsection{Localization}
\label{sec:localization}
In the absence of timing solutions (which generally result in very precise positions for the pulsars), we exploited the hundreds of beams synthesized in each of the seven TRAPUM observations of NGC~1851 to obtain improved localizations for all the newly found pulsars. To do so, we used a custom version of the \texttt{SeeKAT}\footnote{\url{https://github.com/BezuidenhoutMC/SeeKAT}} tied-array beam localization software (Bezuidenhout et al. in prep.). 

For a given observation, \texttt{SeeKAT} takes the S/N values with which a pulsar is detected in the different beams. Using the ratios of the S/N in pairs, it then performs a maximum-likelihood analysis to find the source location that minimises the differences in the ratios, using the beam positions and accounting for their point spread functions.
The latter are calculated with the \texttt{Mosaic}\footnote{\url{https://gitlab.mpifr-bonn.mpg.de/wchen/Beamforming.git}} software \citep{Chen+2021}, using the observation mid-point date and time, as well as the central observing frequency. The custom version of \texttt{SeeKAT} used for this work is also capable of combining information from multiple observations (up to seven, in our case), returning probability contours for the pulsar position that are significantly more accurate than those obtained using a single observation.

%%%%%%%%%%%%%%%%%%%%%%%%%%%%%%%%%%%%%%%%%%%%%%%%%%%%%%%
%   RESULTS
%%%%%%%%%%%%%%%%%%%%%%%%%%%%%%%%%%%%%%%%%%%%%%%%%%%%%%%

\begin{table*}
\caption{The fourteen known pulsars in NGC~1851 and their main characteristics. Pulsar A is the only previously known pulsar and its quoted parameters, except $\langle S_{\rm UHF} \rangle$ and RM, are taken from \citet{Ridolfi+2019}. All the other pulsars are new discoveries presented in this work. Note the three MSPs with massive companions and eccentric orbits. $P$: barycentric spin period; DM: dispersion measure; $\Pb$: orbital period; $\xp$: projected semi-major axis of the pulsar orbit; $e$: eccentricity; $\Mc^{\rm min}$: minimum companion mass as derived from the mass function, assuming a pulsar mass of $\Mp=1.4$~\msun; $\alpha_{\rm J2000}$,~$\delta_{\rm J2000}$: maximum-likelihood right ascension and declination derived with \texttt{SeeKAT} (see Section \ref{sec:localization}); $\langle S_{\rm UHF} \rangle$: average mean flux density in the MeerKAT UHF band (see Section \ref{sec:flux_densities}); RM: rotation measure. The quoted uncertainties on the DMs are the standard error on the average of all the sample measurements, whereas the uncertainties on $\langle S_{\rm UHF} \rangle$ are the standard deviations of all the measurements. The uncertainties on the RMs are those reported by \texttt{rmfit}, as described in Section \ref{sec:polarimetry}. The uncertainties on $\alpha_{\rm J2000}$,~$\delta_{\rm J2000}$ are derived from the 1-$\sigma$ contours returned by \texttt{SeeKAT}.}
\label{tab:discoveries}
\tiny
\centering
\renewcommand{\arraystretch}{1.1}
\vskip 0.1cm
\begin{tabular}{ccrlrrlcllcc}
\hline
\hline
\multicolumn{12}{c}{Summary of the pulsars known in NGC~1851}\\
\hline
Pulsar      & Type        &  \multicolumn{1}{c}{$P$}   & \multicolumn{1}{c}{DM}        &  \multicolumn{1}{c}{$\Pb$}   & \multicolumn{1}{c}{$\xp$}      & \multicolumn{1}{c}{$e$}      & \multicolumn{1}{c}{$\Mc^{\rm min}$}  & \multicolumn{1}{c}{$\alpha_{\rm J2000}$}  & \multicolumn{1}{c}{$\delta_{\rm J2000}$} & $\langle S_{\rm UHF} \rangle$  & RM  \\
        &             & \multicolumn{1}{c}{(ms)}   & \multicolumn{1}{c}{(\dmunit)} & \multicolumn{1}{c}{(d)}   & \multicolumn{1}{c}{(lt-s)}        &          & \multicolumn{1}{c}{(\msun)}    &  &  & ($\mu$Jy) & (\rmunit) \\
\hline
A   &  Binary   & 4.991   & 52.1402(4)   & 18.785      &  36.290     & 0.888       & 0.9     &  05$\h$14$\m$06$\s$.6927(2)  &  $-40\degr02\arcmin48\arcsec.893(2)$         & 241(115) & $-12(4)$ \\ 
B   &  Isolated   & 2.816   & 52.076(3)   & --      &  --     & --       & --    & 05$\h$14$\m$06$\s$.73(8)      & $-40\degr02\arcmin47\arcsec.9(6)$         & 82(40) & $-13(4)$  \\ 
C   &  Isolated   & 5.565   & 52.053(5)    & --      &  --     & --       & --    & 05$\h$14$\m$06$\s$.69(4)      & $-40\degr02\arcmin49\arcsec.4(3)$        & 76(31)  & $-8(5)$\\ 
D   &  Binary     & 4.554   & 52.219(5)   & 5.686   &  9.800  & 0.862    & 0.48    & 05$\h$14$\m$06$\s$.73(5)      & $-40\degr02\arcmin48\arcsec.2(5)$      & 117(57)  & $-12(4)$\\ 
E   &  Binary     & 5.595   & 51.930(3)   & 7.448   &  27.826 & 0.708    & 1.53    & 05$\h$14$\m$06$\s$.68(6)       & $-40\degr02\arcmin47\arcsec.7(5)$      & 65(17)  & $-6(4)$  \\ 
F   &  Binary     & 4.329   & 51.620(9)    & 4.313   &  4.073  &  0     &  0.23    & 05$\h$14$\m$06$\s$.7(1)       & $-40\degr02\arcmin48\arcsec.5(8)$         & 26(14) & -- \\ 
G   &  Binary     & 3.803   & 50.959(3)   & 2.636   &  1.997  & 0       & 0.14    & 05$\h$14$\m$06$\s$.63(5)      & $-40\degr02\arcmin48\arcsec.1(4)$       & 99(23) & $-12(3)$  \\ 
H   &  Binary     & 5.506   & 52.226(6)    & 16.944  &  12.805 & 0.020        & 0.28    & 05$\h$14$\m$06$\s$.6(1)  & $-40\degr02\arcmin48\arcsec(1)$       & 32(18) & $-11(6)$  \\ 
I   &  Binary     & 32.654  & 52.58(3)    & 0.985   &  1.090  & 0.0098   & 0.15    & 05$\h$14$\m$06$\s$.75(7)      & $-40\degr02\arcmin48\arcsec.2(6)$       & 30(7) & -- \\ 
J   &  Isolated   & 6.633   & 52.017(7)    & --      &  --      & --       & --    & 05$\h$14$\m$06$\s$.7(2)    & $-40\degr02\arcmin48\arcsec(2)$        & 26(17) & $-14(6)$ \\ 
K   &  Isolated   & 4.692   & 51.928(2)   & --      &  --      & --       & --    & 05$\h$14$\m$06$\s$.8(2)    & $-40\degr02\arcmin48\arcsec(2)$   & 29(5)   & $-13(6)$     \\ 
L   &  Binary     & 2.959   & 51.245(3)   & 1.141   &  1.326   & 0        & 0.17    & 05$\h$14$\m$06$\s$.8(1)       & $-40\degr02\arcmin48\arcsec.0(8)$  & 35(10)   & --    \\ 
M   &  Isolated   & 4.798   & 51.678(14)  & --      &  --      & --       & --    & 05$\h$14$\m$06$\s$.76$^{\ddag}$    & $-40\degr02\arcmin47\arcsec.60^{\ddag}$      & 26(17) & $-13(5)$\\ 
N   &  Isolated   & 5.568   & 51.068(4)    & --      &  --      & --       & --    & 05$\h$14$\m$06$\s$.7(1)     & $-40\degr02\arcmin47\arcsec(1)$     & 31(6) & -- \\ 
\hline
\multicolumn{12}{l}{$^{\ddag}$ Localization of pulsar M with \texttt{SeeKAT} was not possible due to the faintness and scarcity of its detections. Therefore, the reported coordinates are  }\\
\multicolumn{12}{l}{\phantom{$^{\ddag}$} those of the central beam of the TRAPUM tilings, where pulsar M is always detected with highest S/N.}\\
\end{tabular}
\end{table*}

\begin{figure*}
	\includegraphics[width=0.23\textwidth]{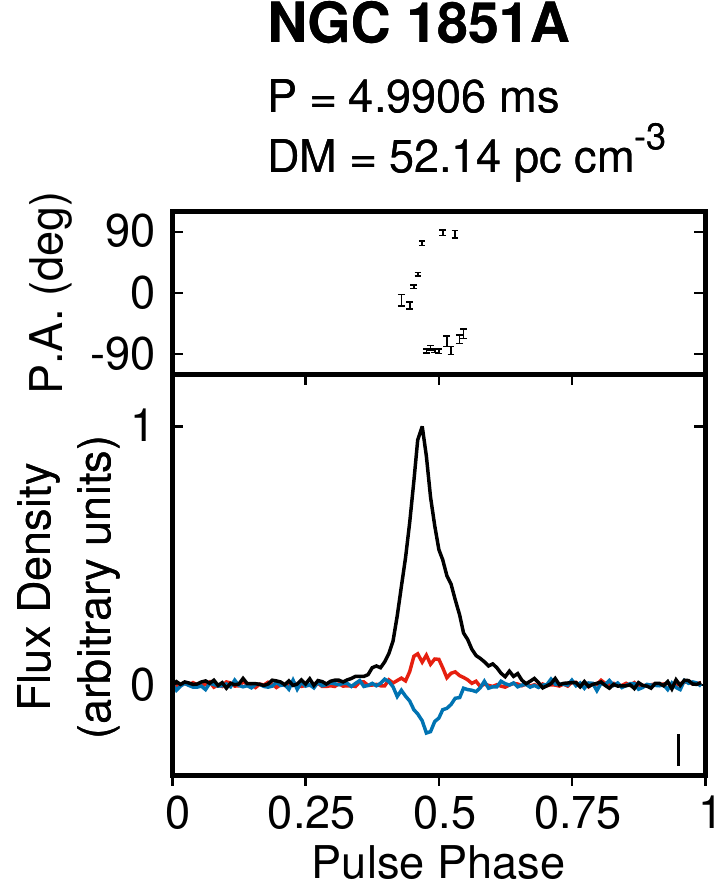}\quad \,
	\includegraphics[width=0.23\textwidth]{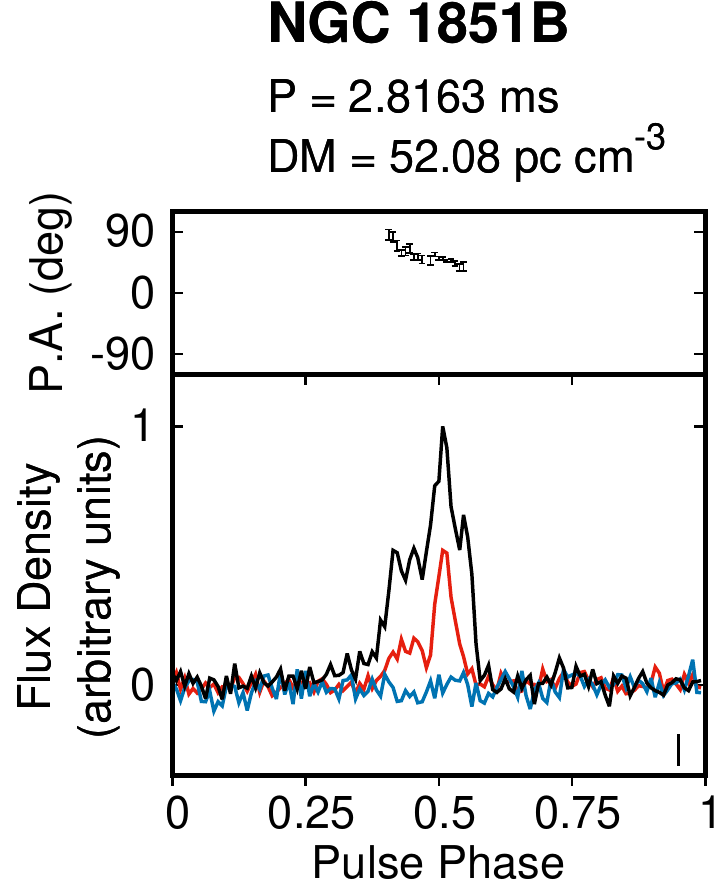}\quad \,
	\includegraphics[width=0.23\textwidth]{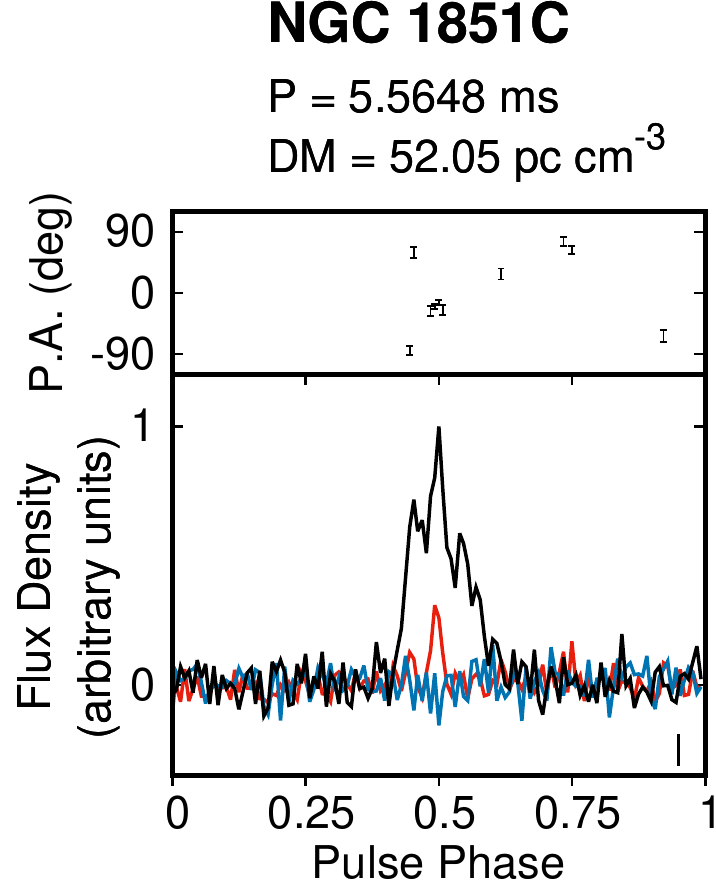}\quad \,
	\includegraphics[width=0.23\textwidth]{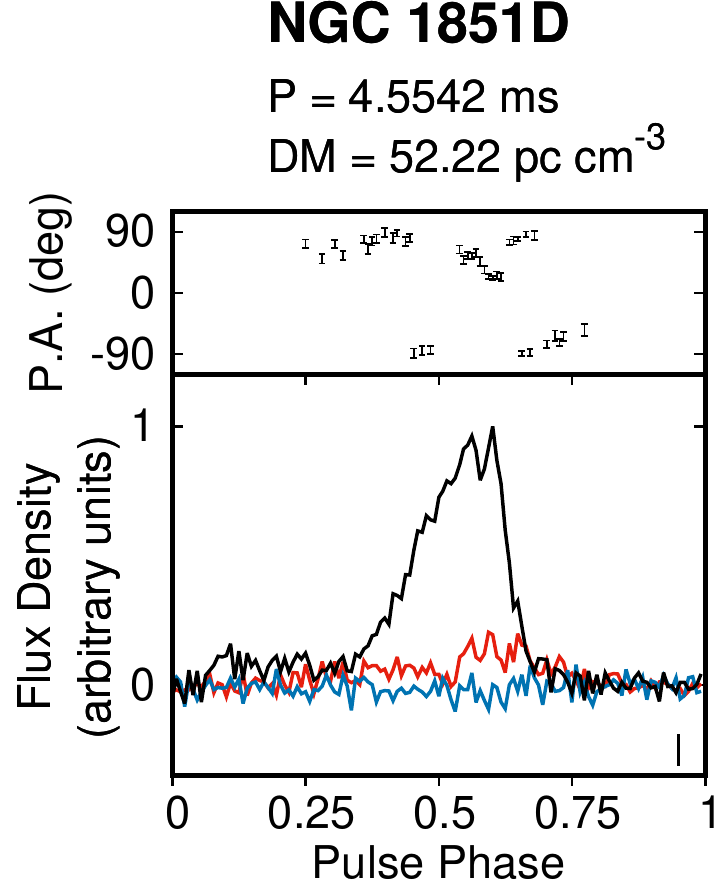}
	\vskip 0.35 cm
	\includegraphics[width=0.23\textwidth]{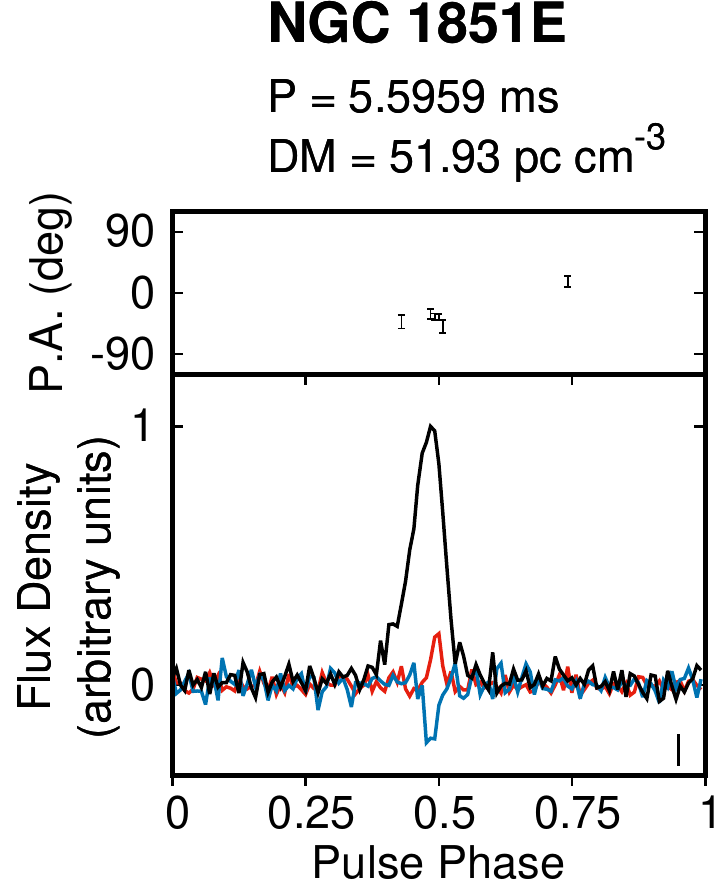}\quad \,
	\includegraphics[width=0.23\textwidth]{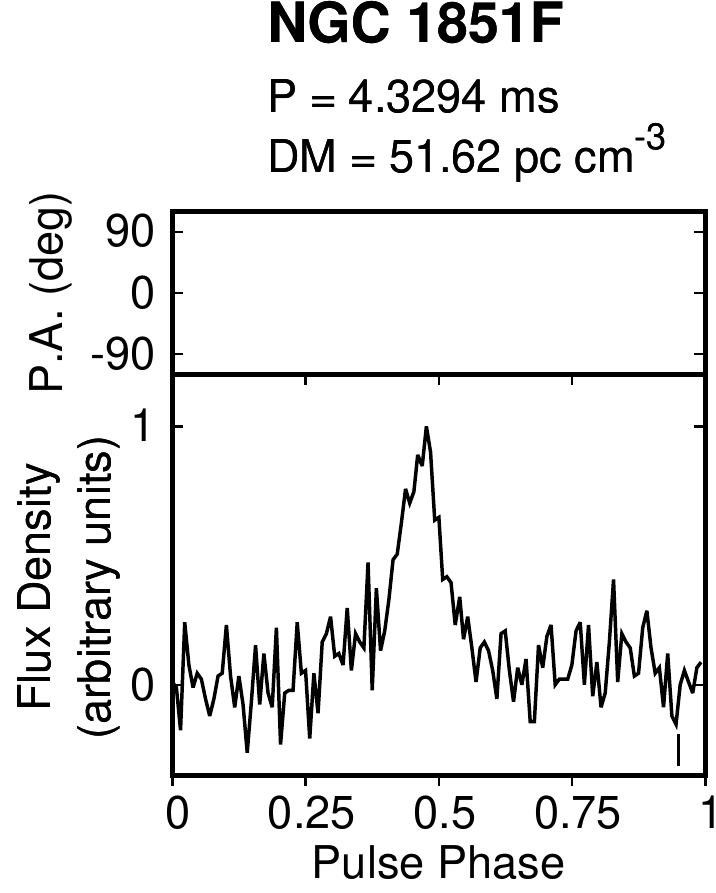}\quad \,
	\includegraphics[width=0.23\textwidth]{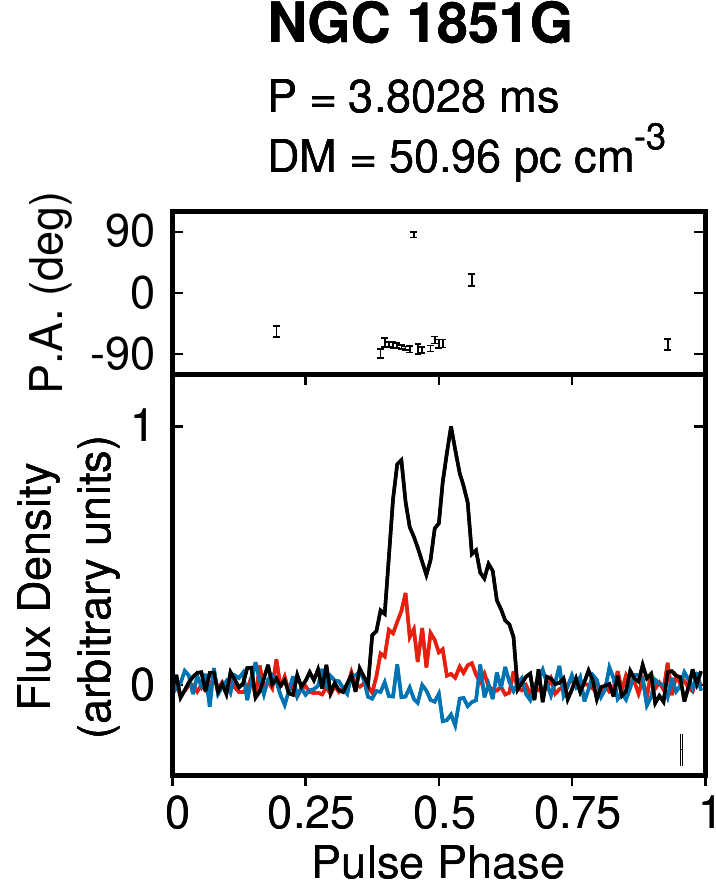}\quad \,
	\includegraphics[width=0.23\textwidth]{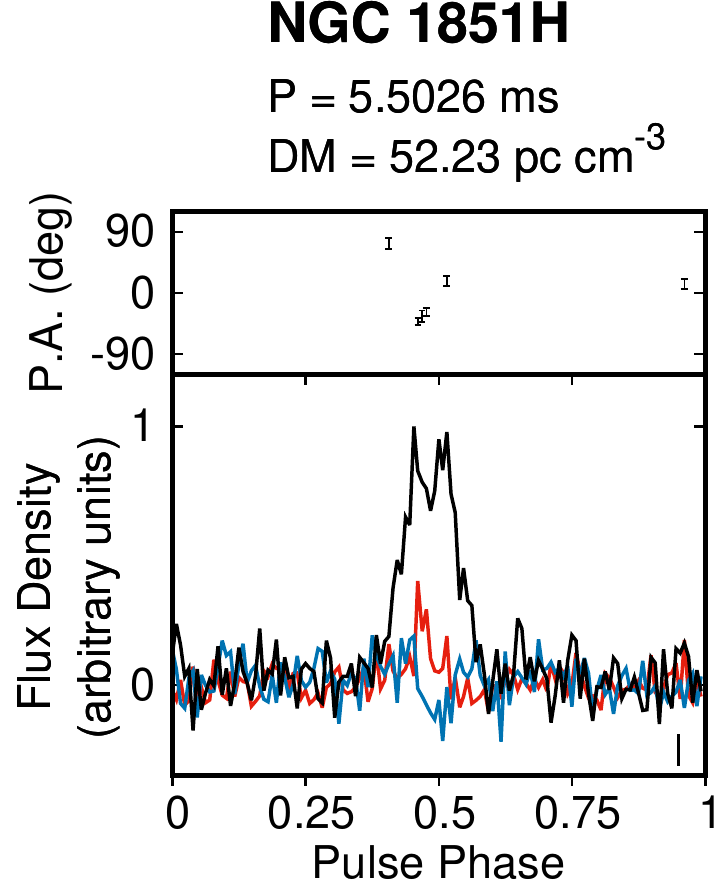}
	\vskip 0.35 cm
	\includegraphics[width=0.23\textwidth]{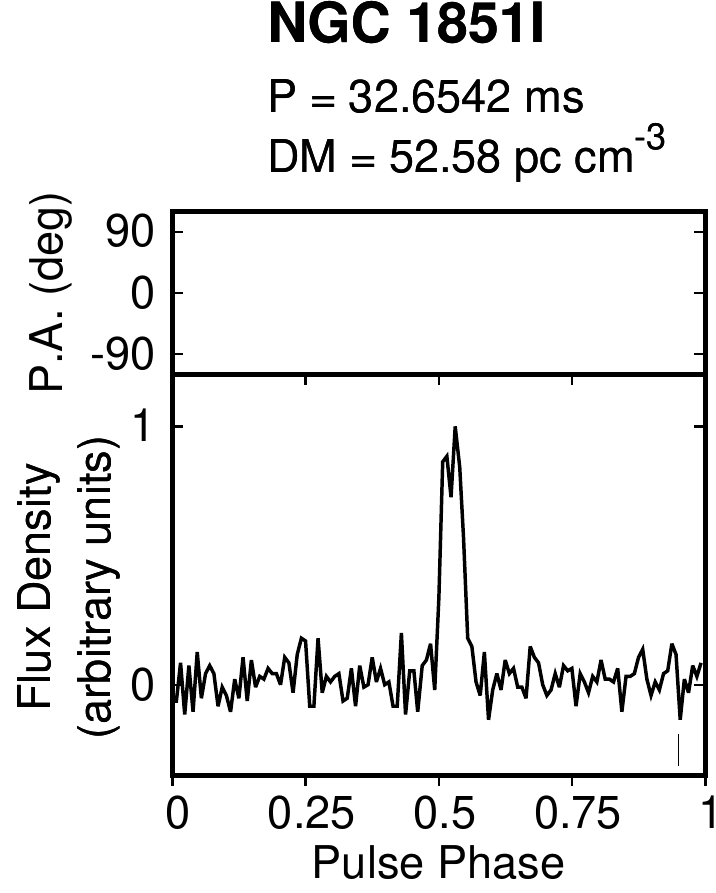}\quad \,
	\includegraphics[width=0.23\textwidth]{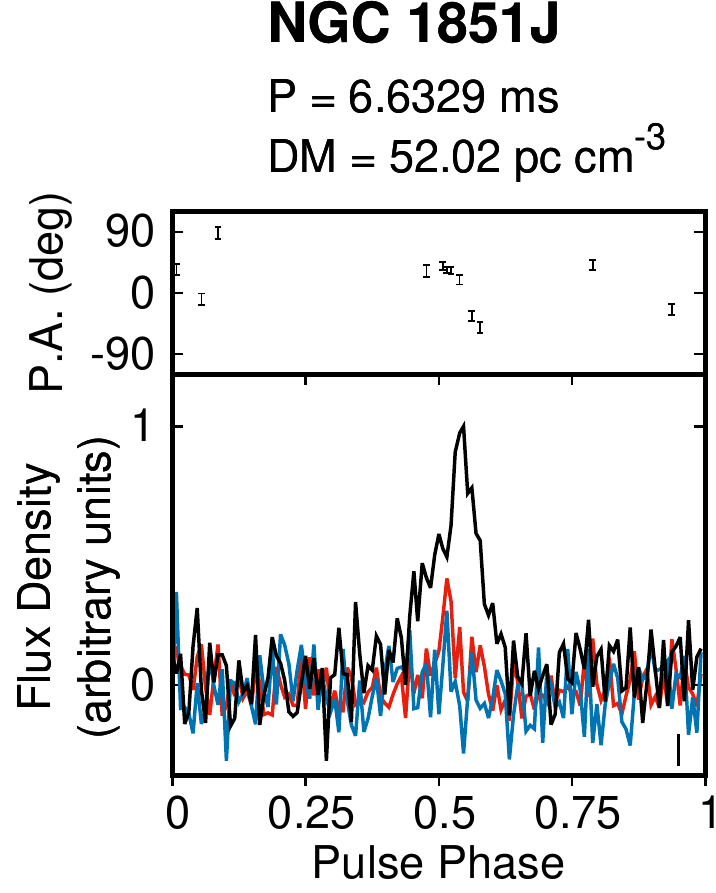}\quad \,
	\includegraphics[width=0.23\textwidth]{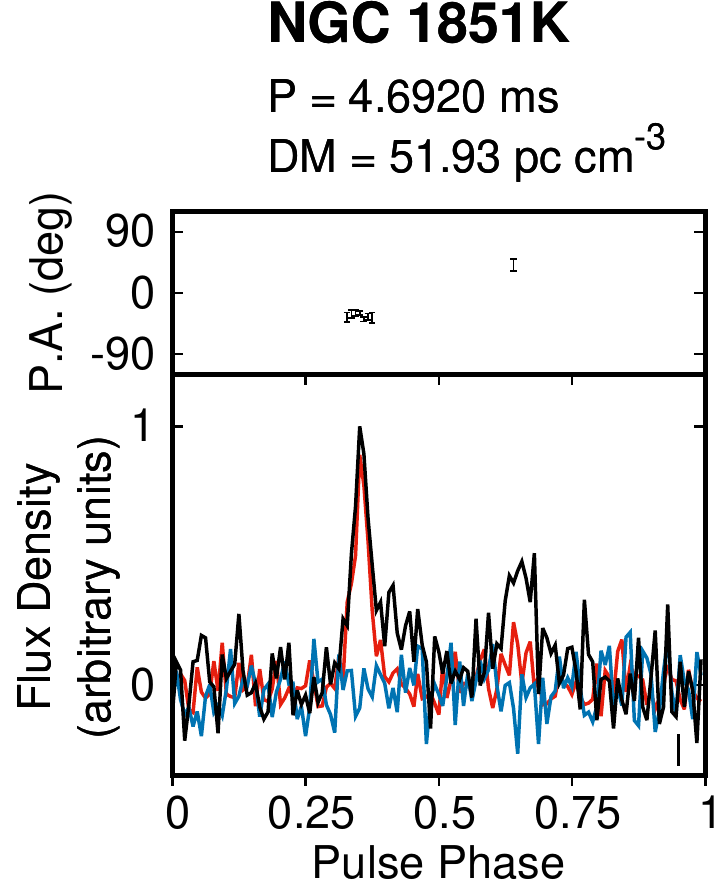}\quad \,
	\includegraphics[width=0.23\textwidth]{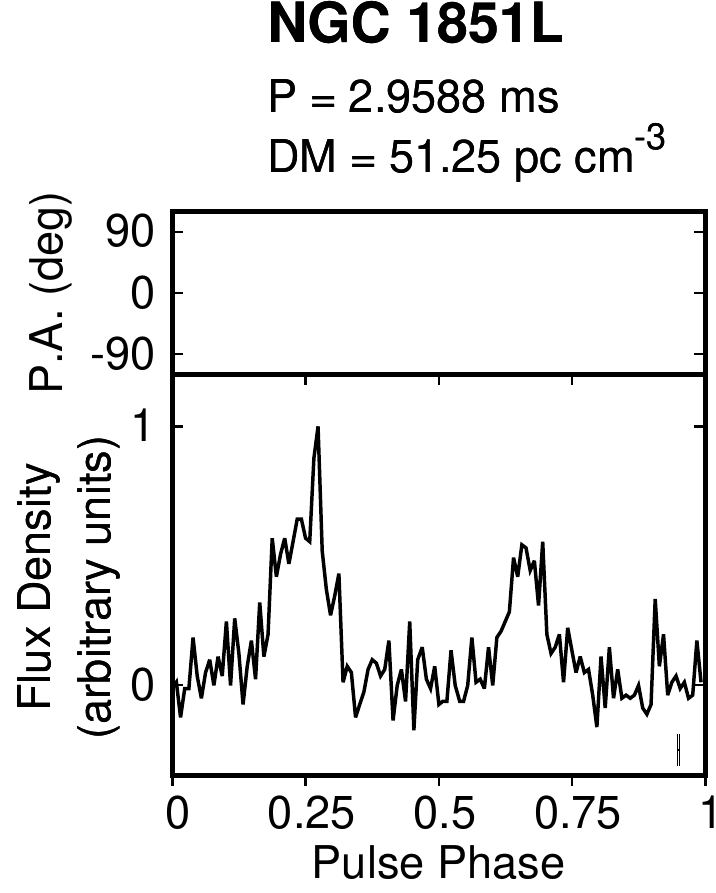}
	\vskip 0.35 cm
	\includegraphics[width=0.23\textwidth]{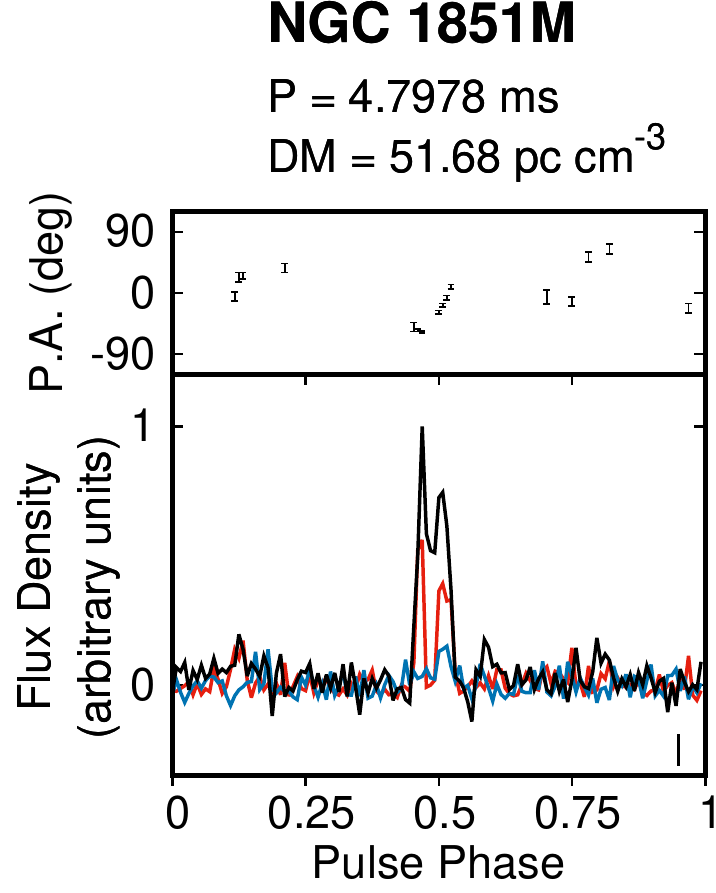}\quad \,
	\includegraphics[width=0.23\textwidth]{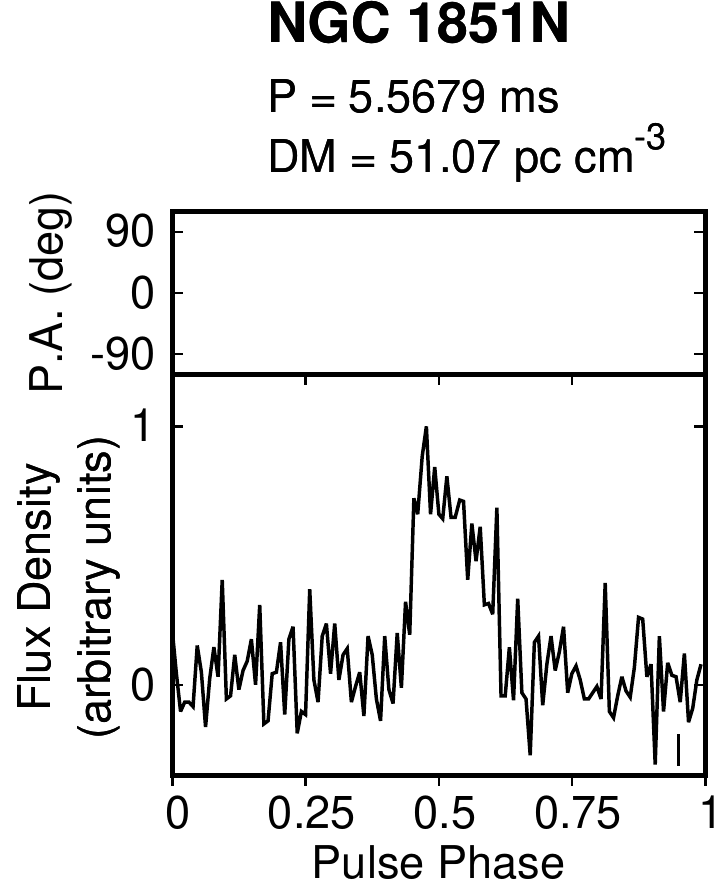}
	
  	\caption{High-S/N integrated pulse profiles of the 14 pulsars in NGC~1851, each obtained summing the brightest detections available in the MeerKAT UHF band. \emph{Bottom panels:} total intensity (black line) and, for the pulsars where some significant polarization was measurable, linear (red line) and circular (blue line) polarization profiles. \emph{Top panels:} position angle (P.A.) of the linear polarization, showing measurements that are at least 3-$\sigma$ significant. We follow the PSR/IEEE convention \citep{vanStraten+2010}, where the left-hand circular polarization is positive and the P.A. angle increases counterclockwise on the sky. The horizontal section of the short bars on the bottom-right of each panel shows the effective time resolution, which accounts for the sampling time (9.41 $\mu$s) and the intra-channel smearing; in all cases it is $<14\,\mu$s.}
  	\label{fig:integrated_profiles}
\end{figure*}

\begin{figure*}
\centering
	\includegraphics[width=0.240\textwidth]{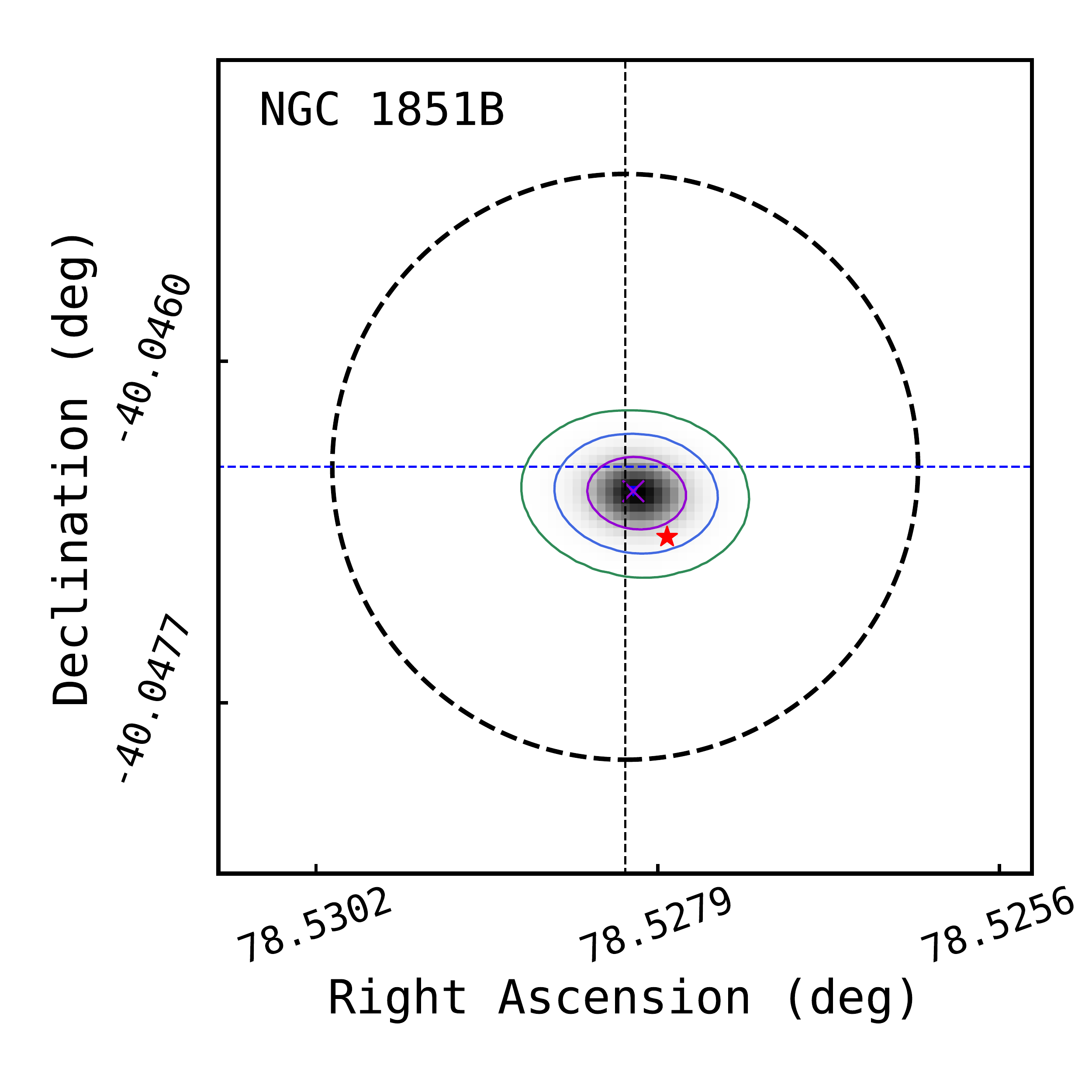}
	\,
	\includegraphics[width=0.240\textwidth]{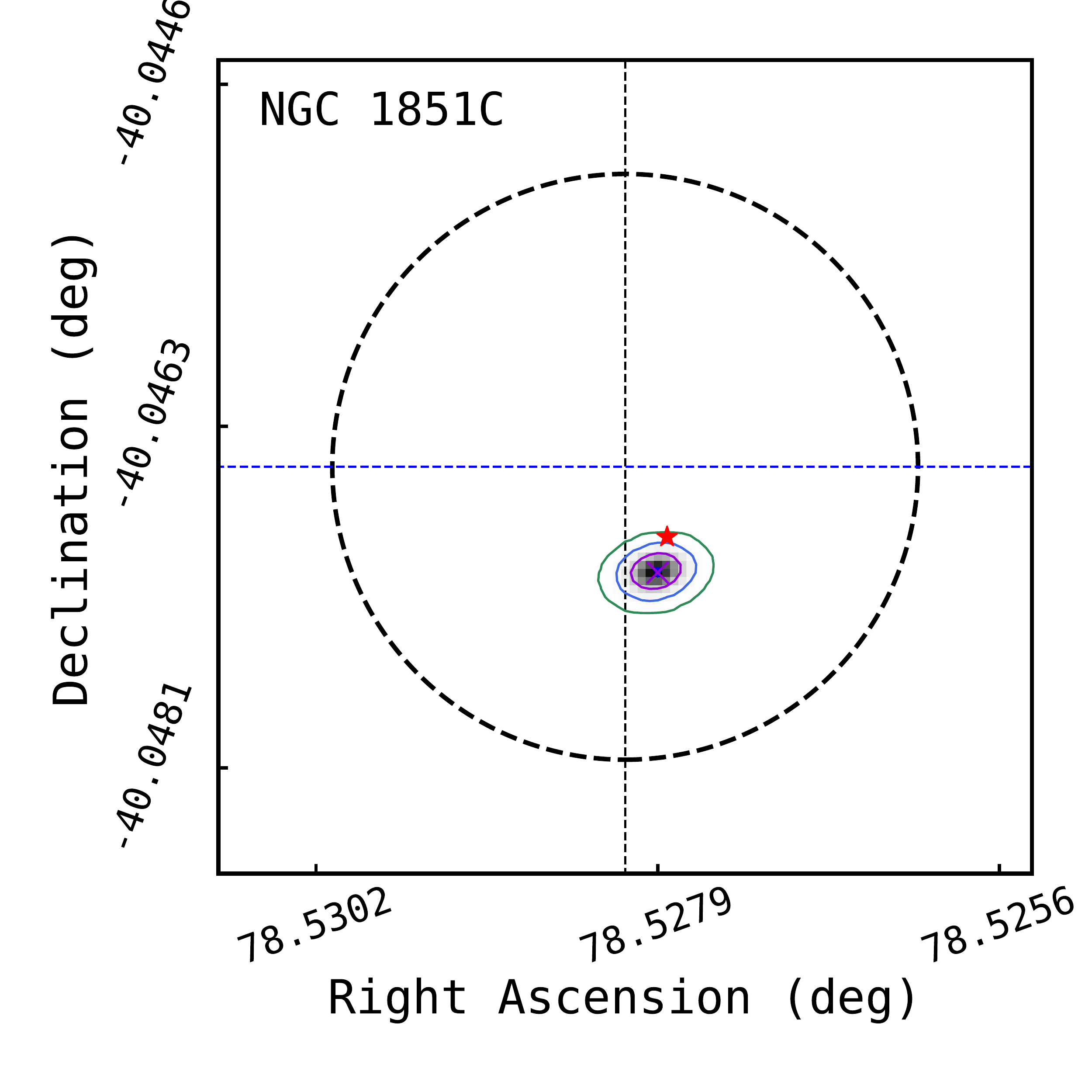}
	\,
	\includegraphics[width=0.240\textwidth]{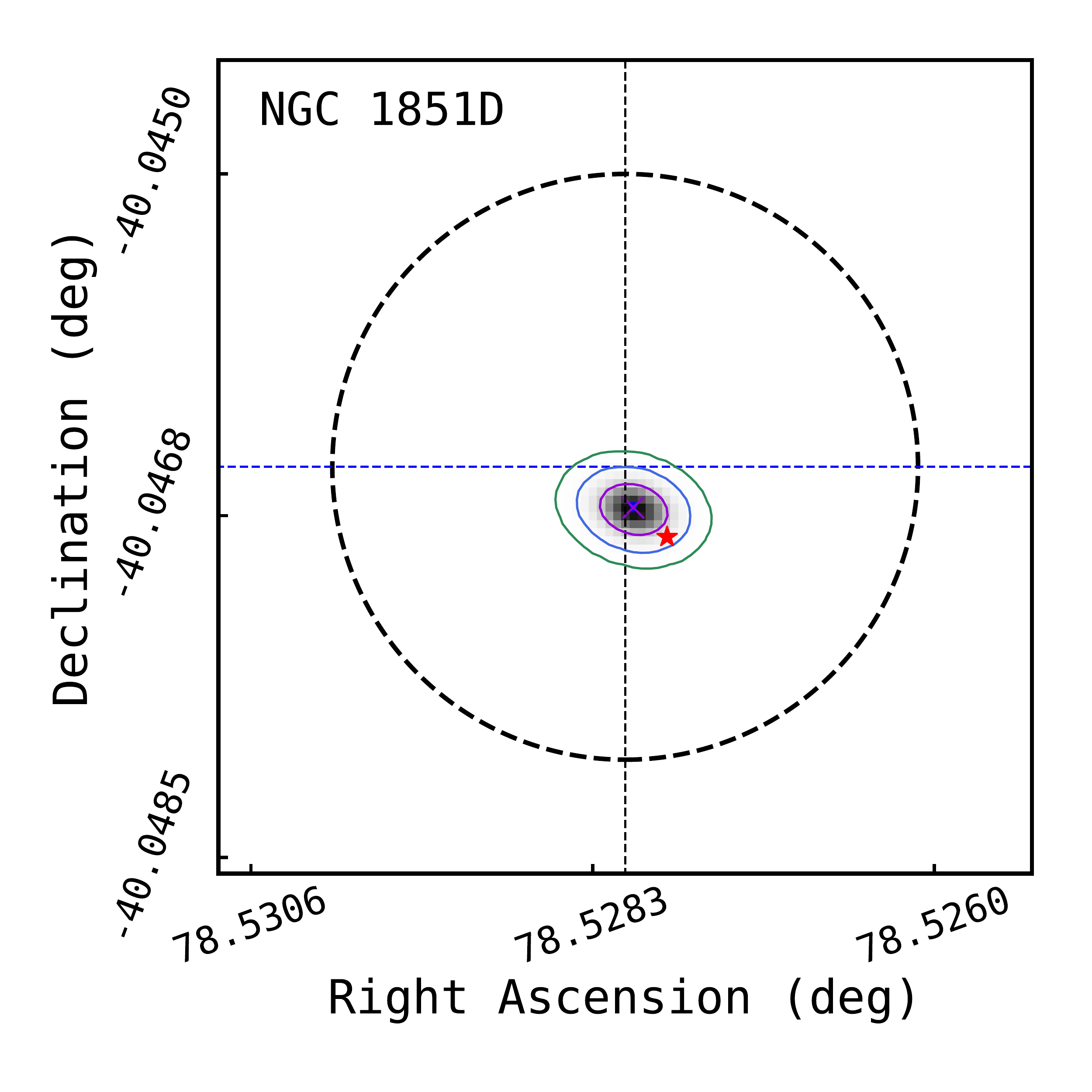}
	\,
	\includegraphics[width=0.240\textwidth]{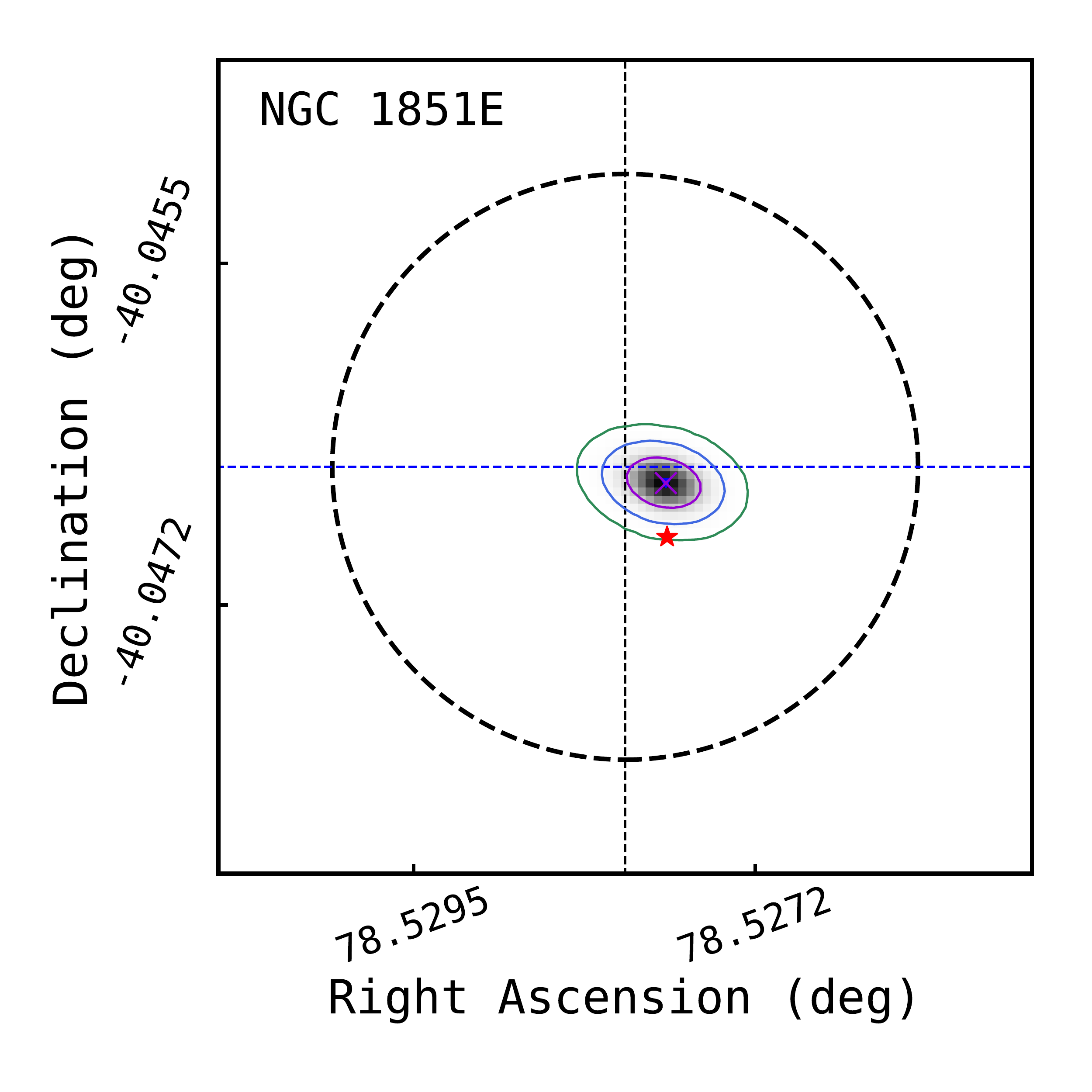}
	\,
	\newline
		\includegraphics[width=0.240\textwidth]{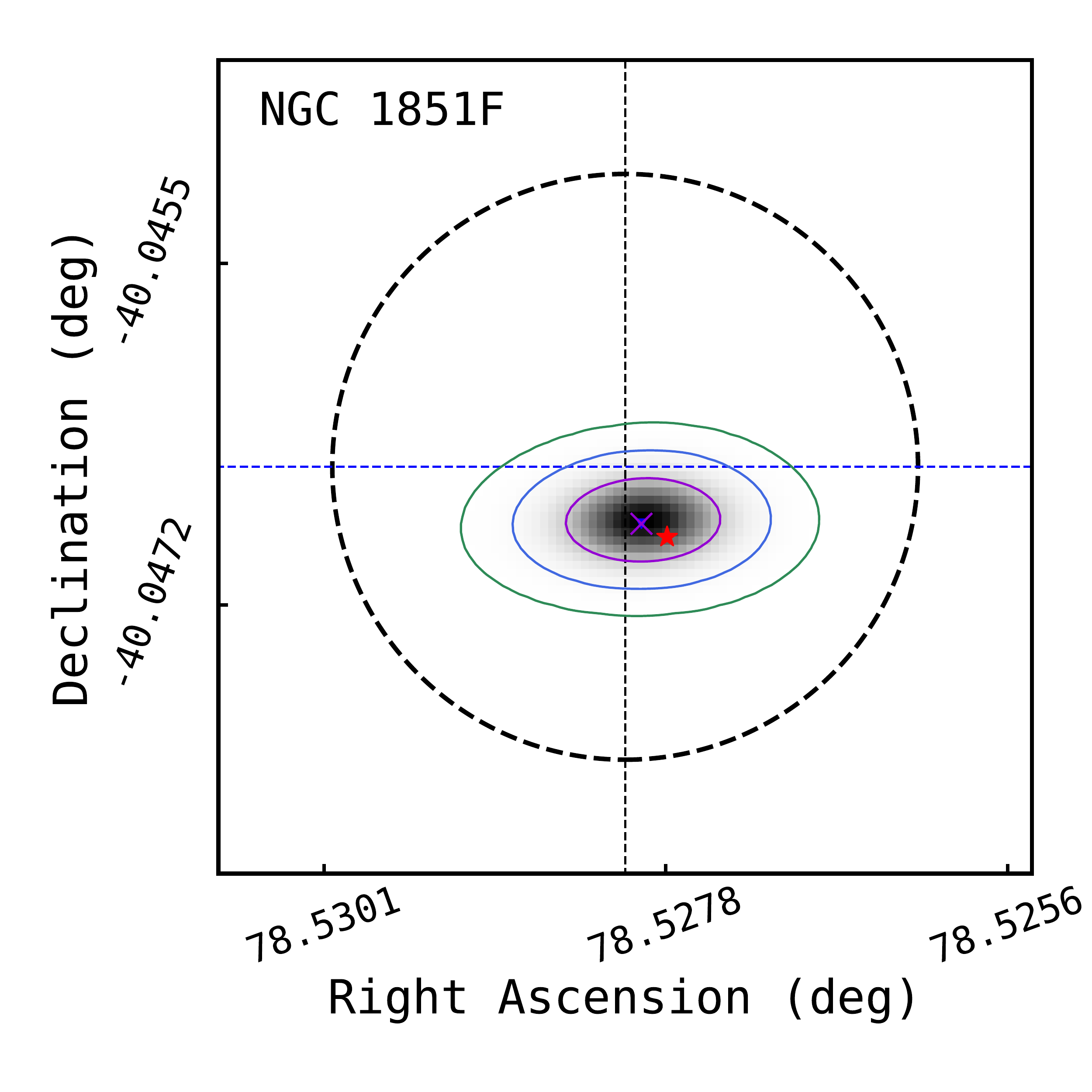}
	\,
	\includegraphics[width=0.240\textwidth]{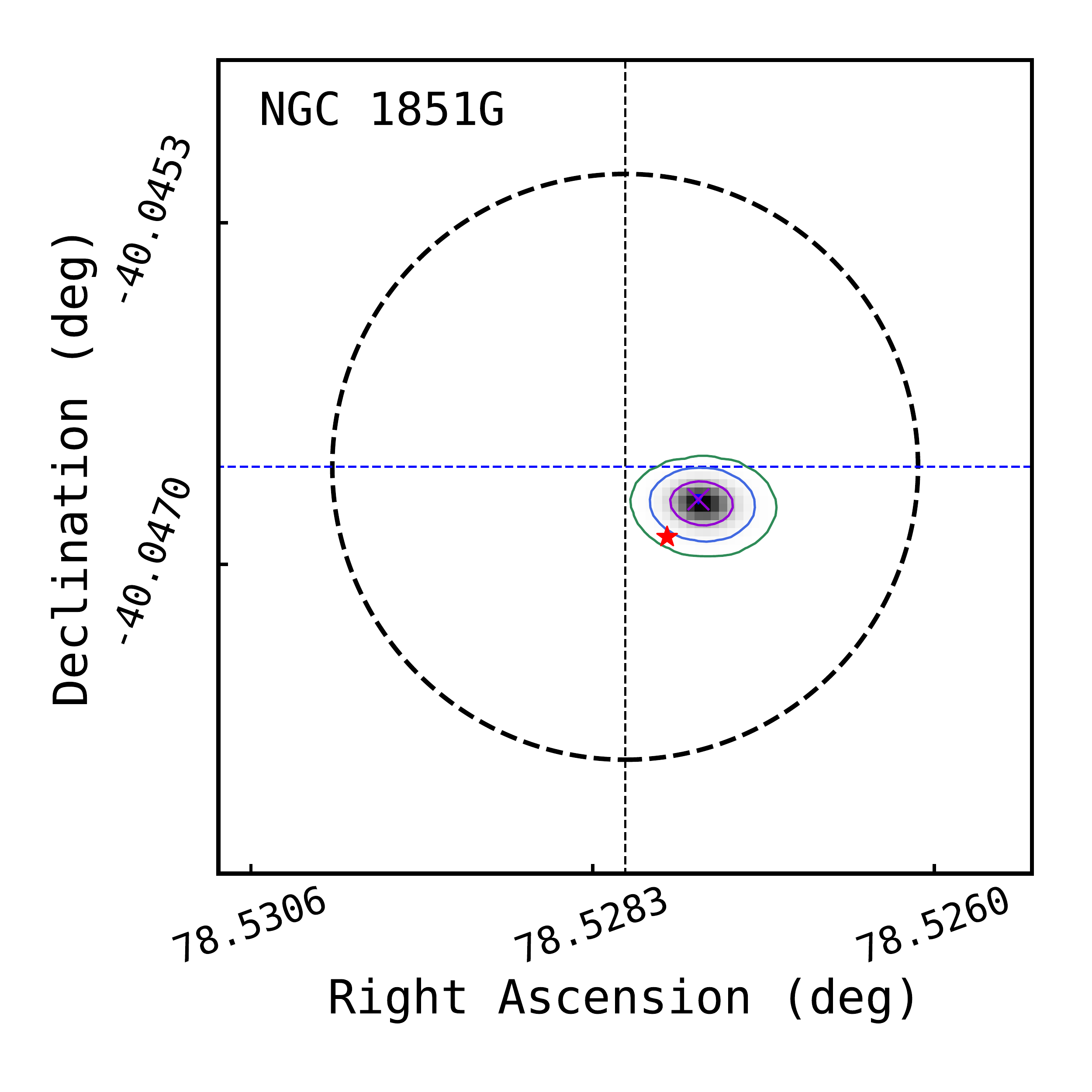}
	\,
	\includegraphics[width=0.240\textwidth]{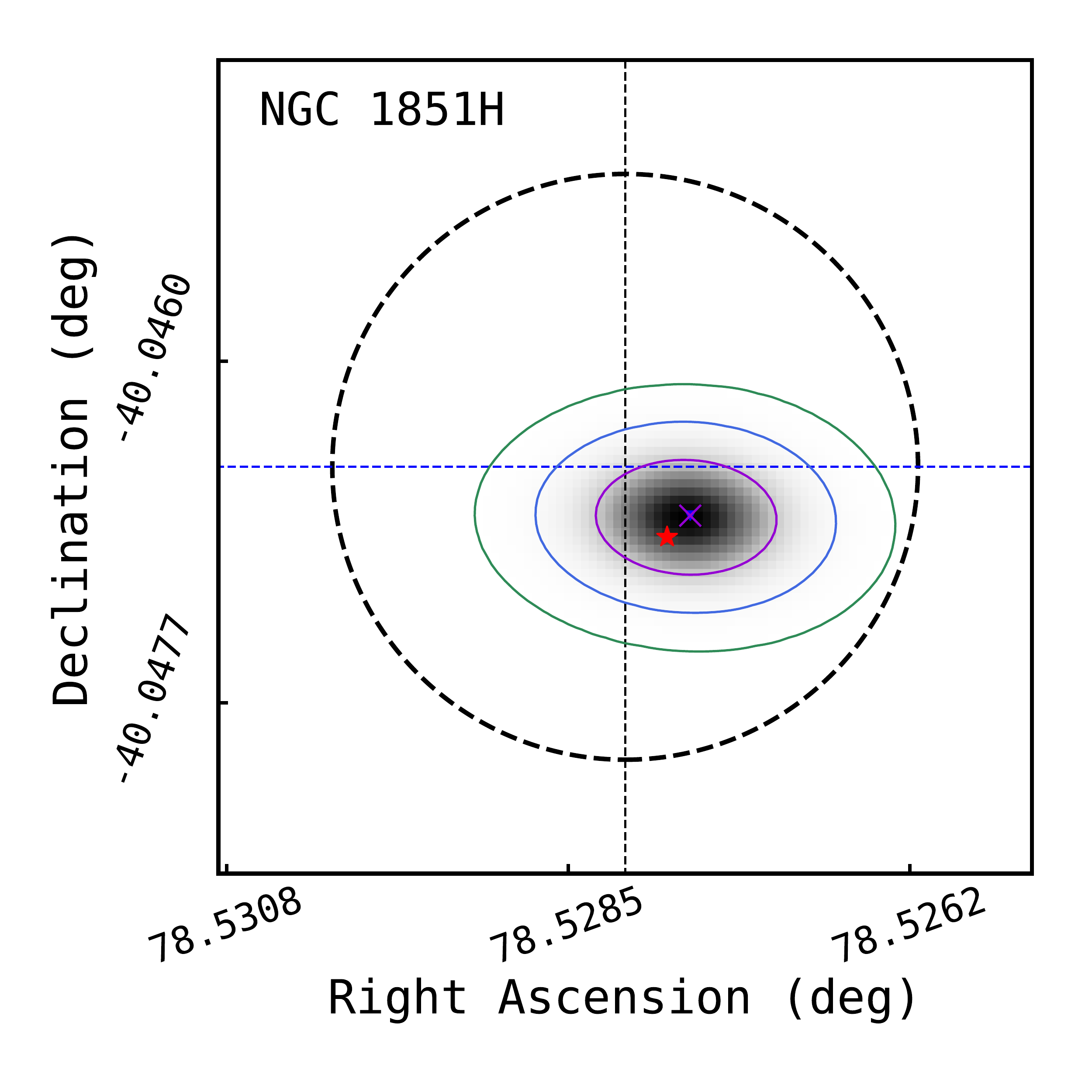}
	\,
	\includegraphics[width=0.240\textwidth]{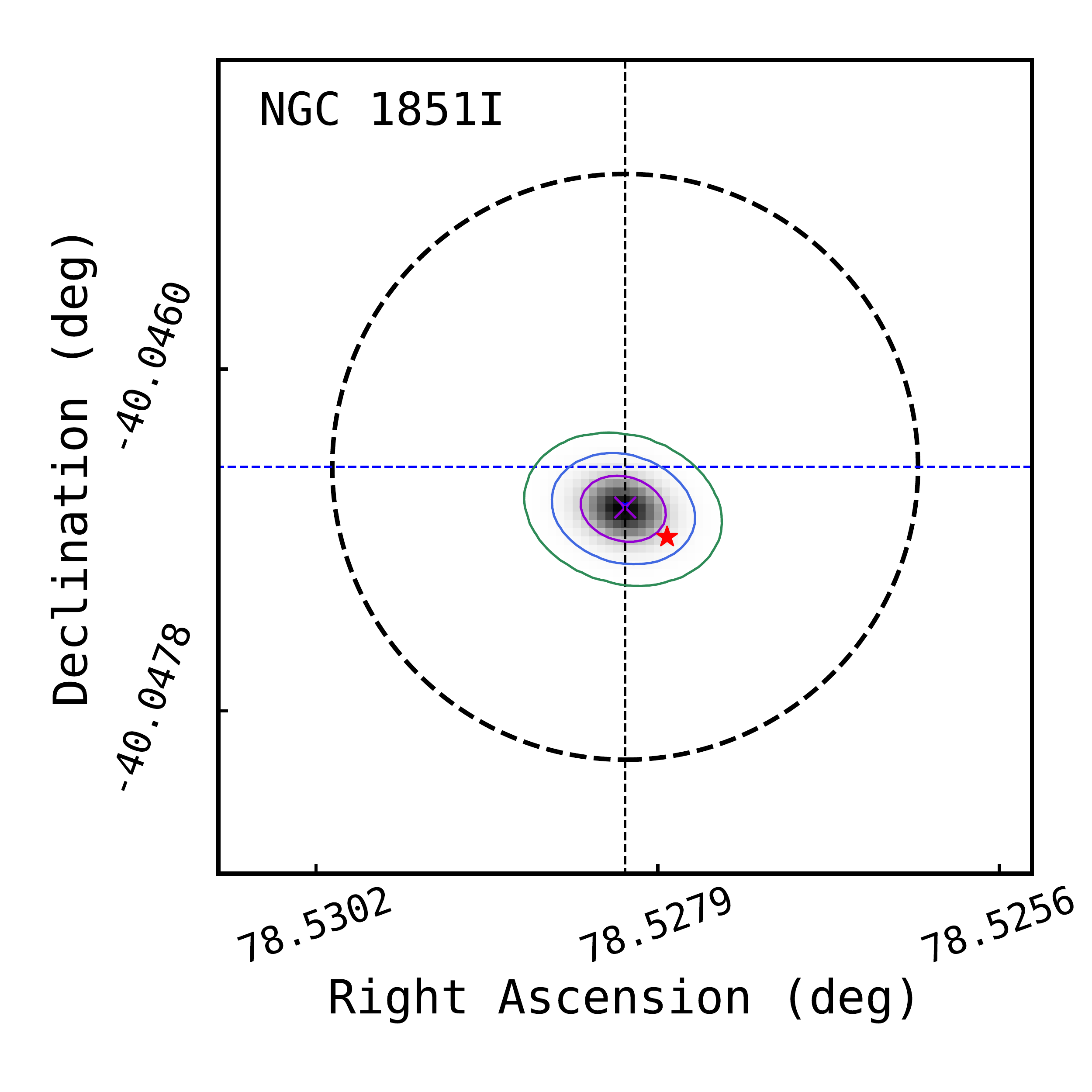}
	\,
		\newline
		\includegraphics[width=0.240\textwidth]{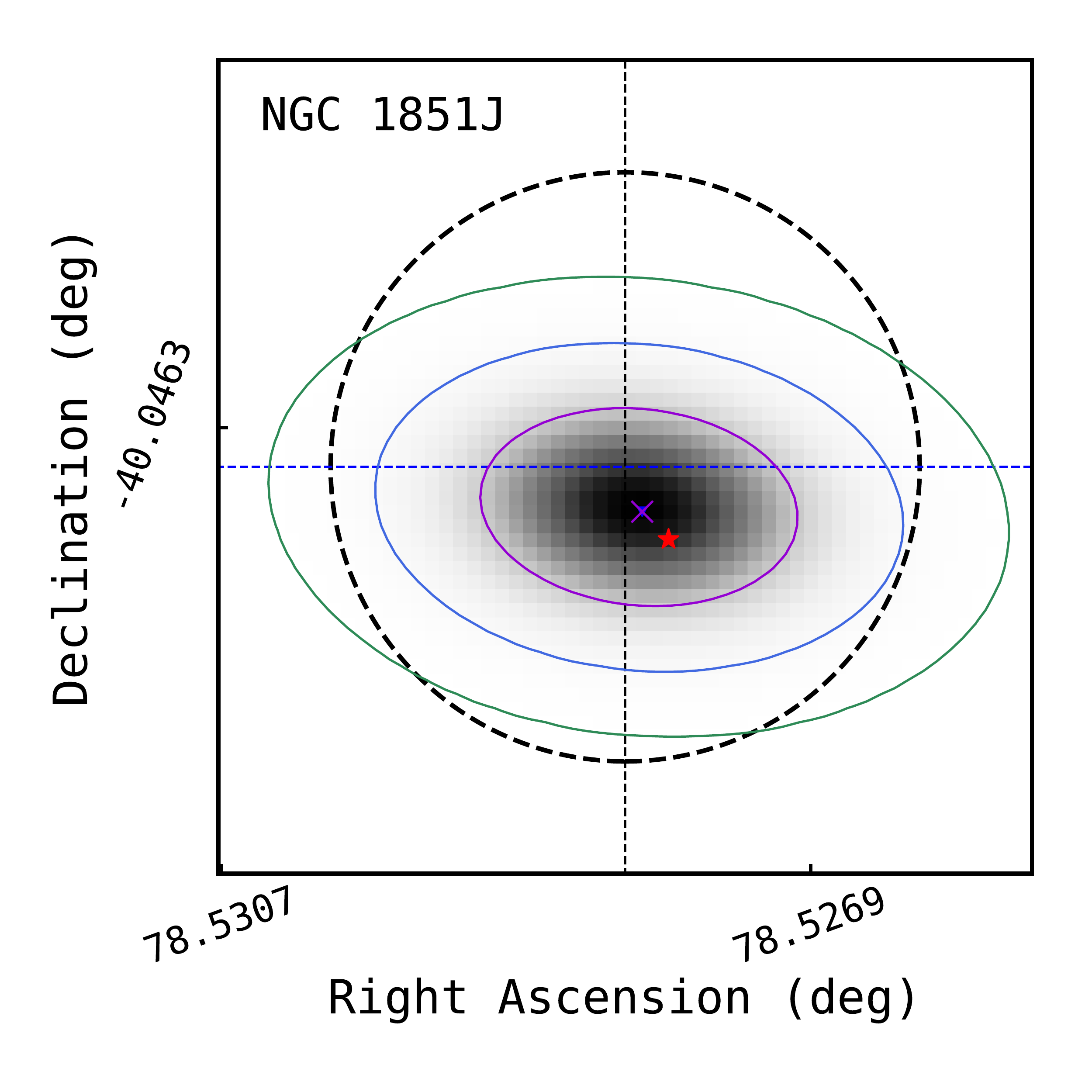}
	\,
	\includegraphics[width=0.240\textwidth]{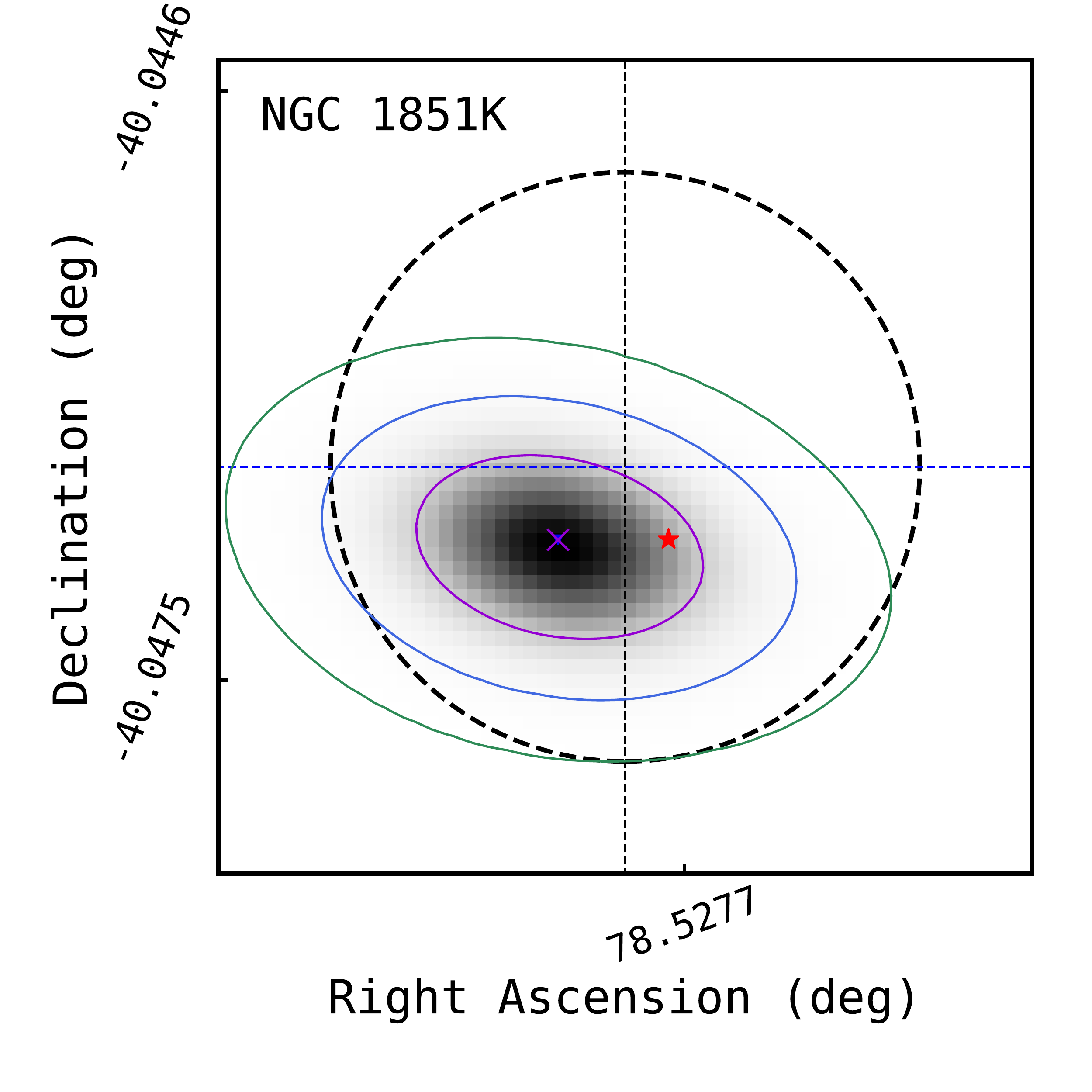}
	\,
	\includegraphics[width=0.240\textwidth]{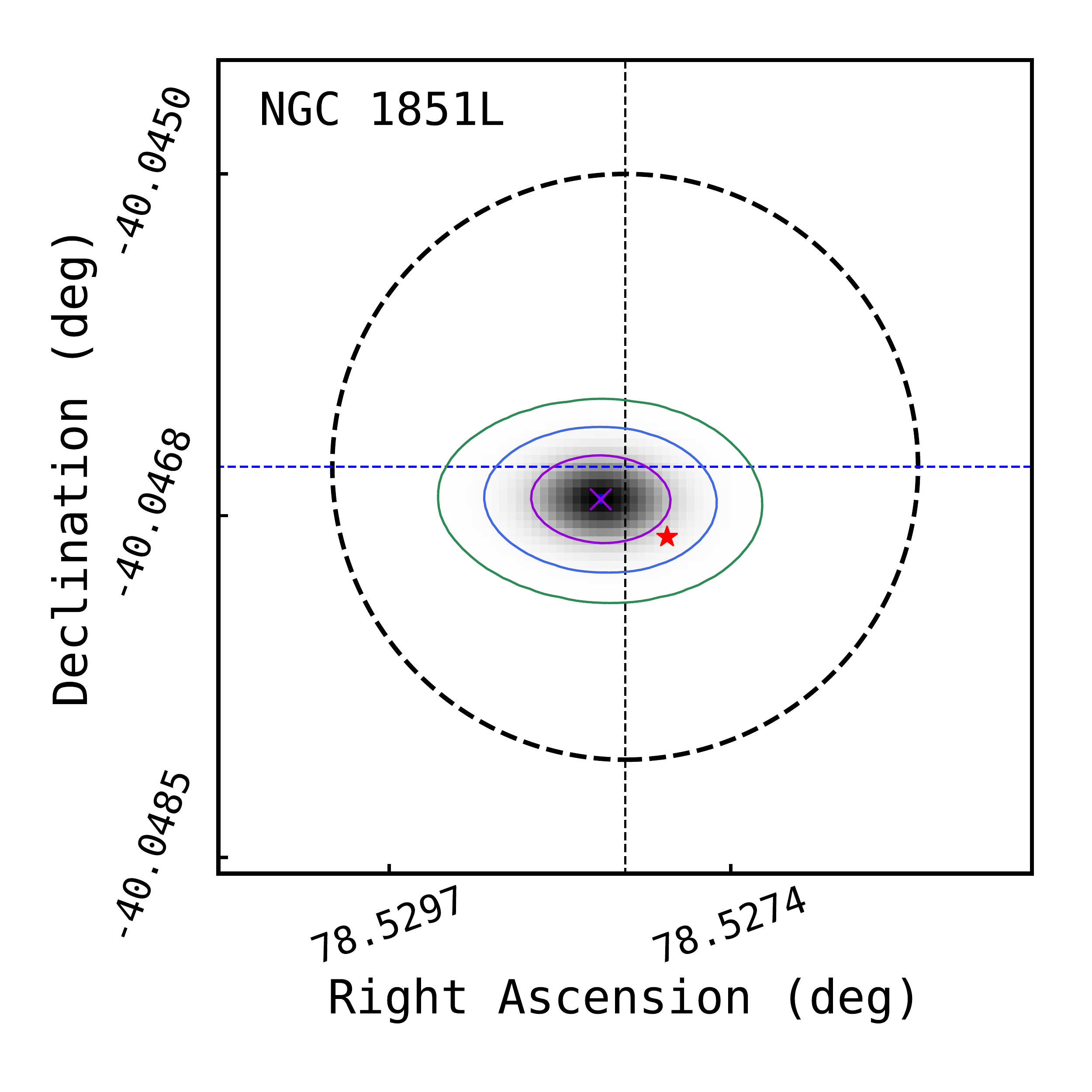}
	\,
	\includegraphics[width=0.240\textwidth]{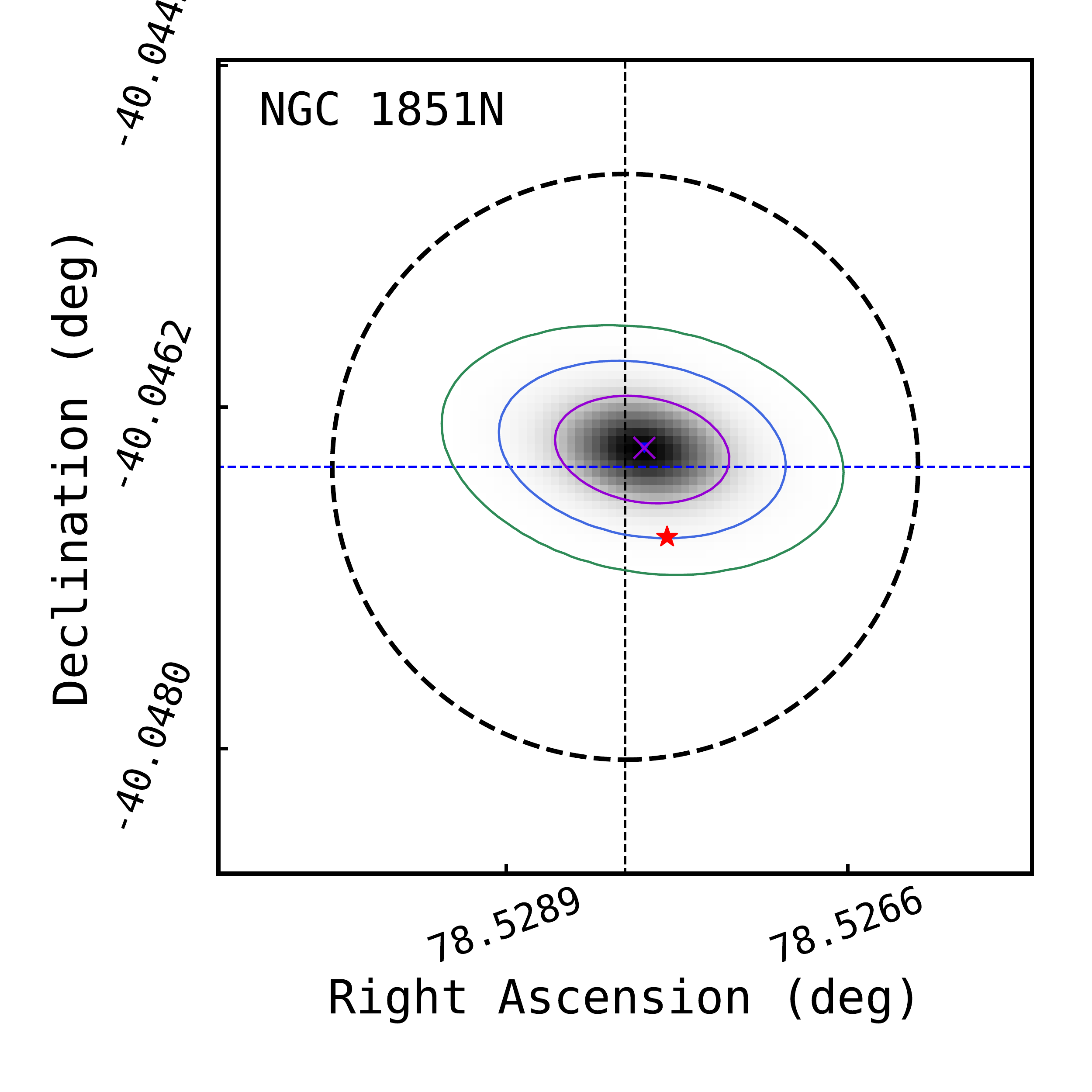}
	\,

  	\caption{Positions of the new pulsars determined by \texttt{SeeKAT}, relative to the centre of the cluster (the intersection of the horizontal and vertical dashed lines) and the position of NGC~1851A, indicated by the red star. The grey scale background indicates probability density, while the purple, blue and green ellipses represent the 68\%, 95\% and 99.7\% probability contours, respectively. The black dashed circle indicates the core radius of NGC~1851, located at 0.09 arcmin from its nominal center \citep{Harris2010}. The detections of NGC~1851M were too faint to obtain a reasonable position estimate with \texttt{SeeKAT}. All pulsars are detected with highest S/N in the central beams of all TRAPUM observations.}
  	\label{fig:localizations}
\end{figure*}

\begin{figure*}
\centering
	\includegraphics[width=0.45\textwidth]{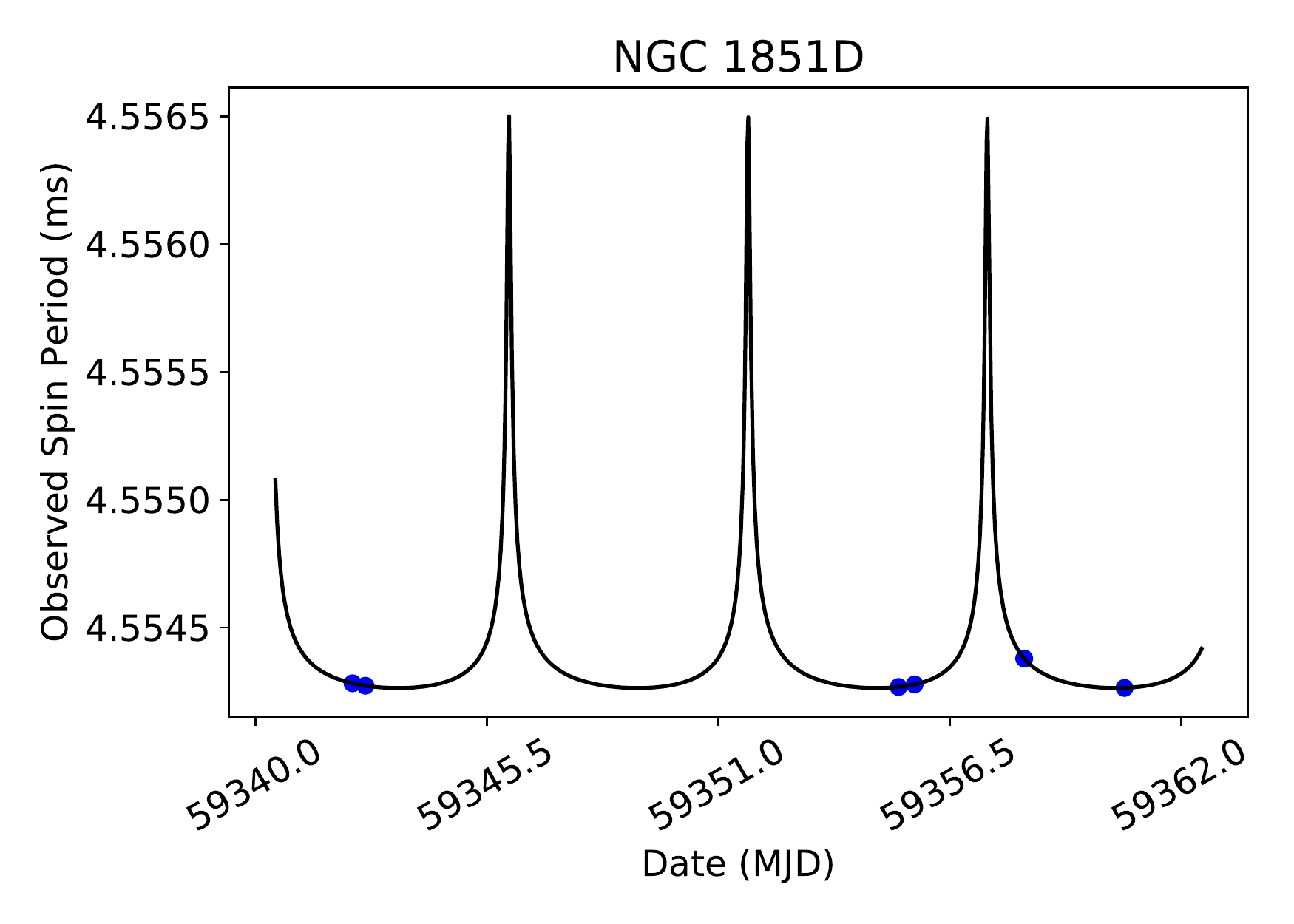}
	\qquad
	\includegraphics[width=0.45\textwidth]{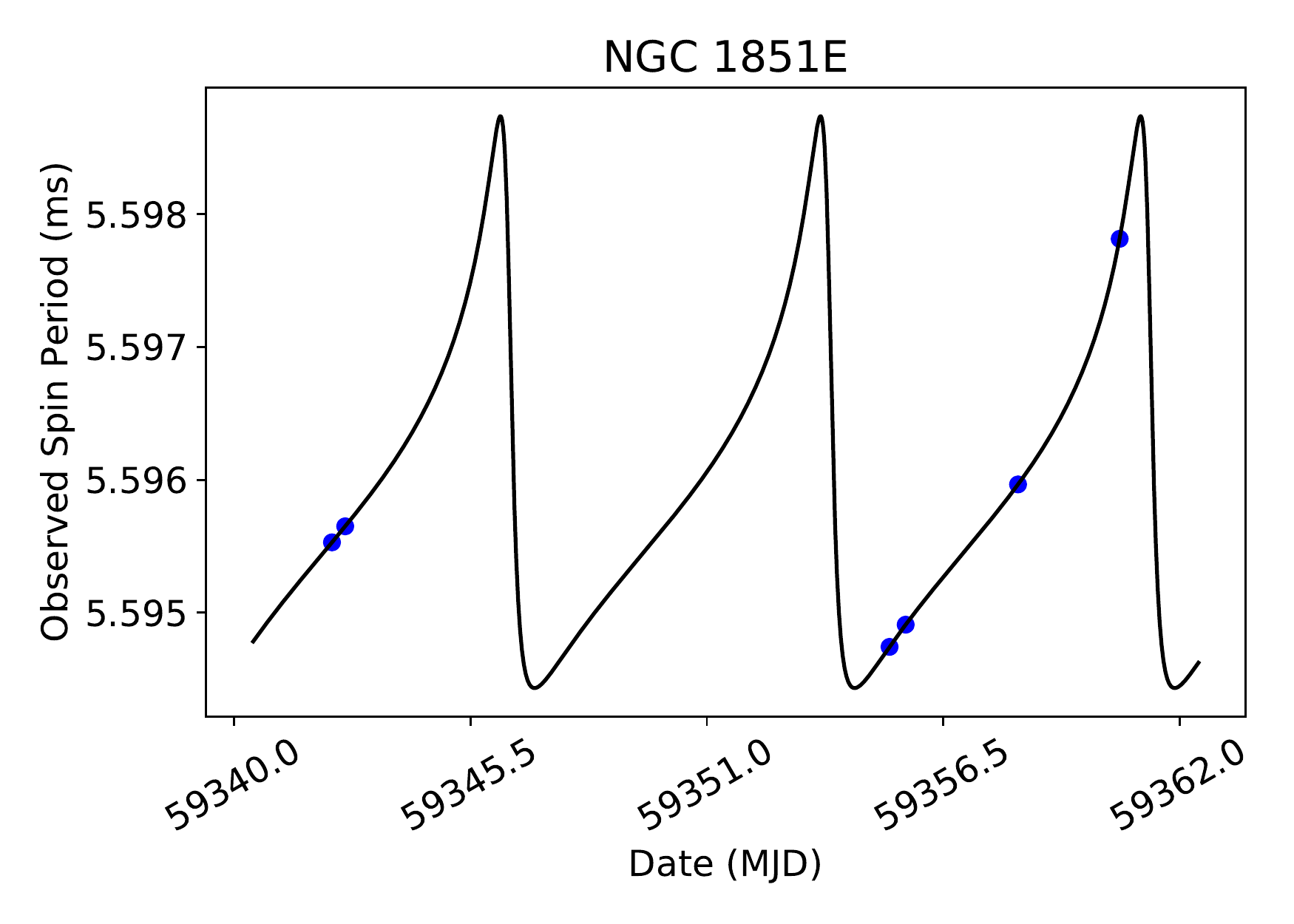}
  	\caption{Spin period as a function of time as observed in May 2021 for the two eccentric binary pulsars NGC~1851D (left) and NGC~1851E (right). The former is in a 5.69-d orbit with an eccentricity of 0.86. The latter is in a 7.44-d orbit with an eccentricity of 0.71.}
  	\label{fig:NGC1851DE_Pobs_vs_MJD}
\end{figure*}

\begin{figure}
\centering
	\includegraphics[width=0.42\columnwidth]{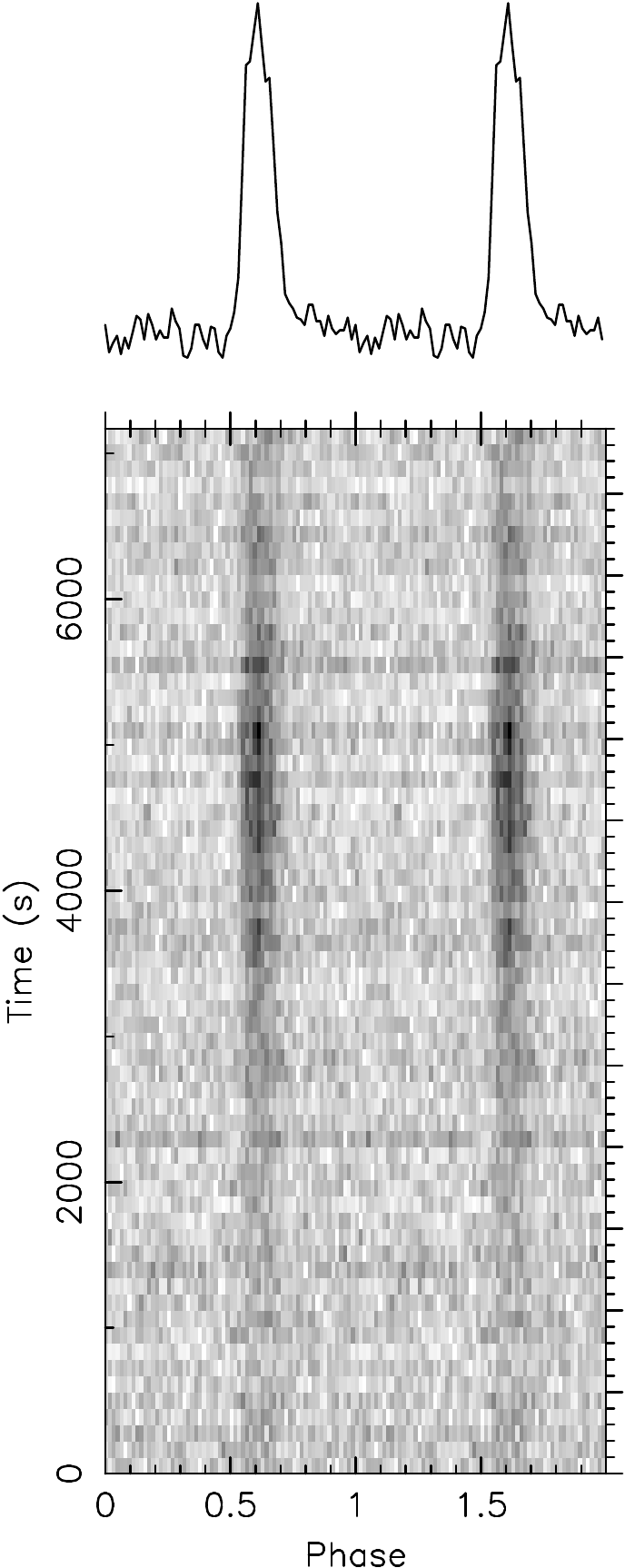}
	\qquad
	\includegraphics[width=0.42\columnwidth]{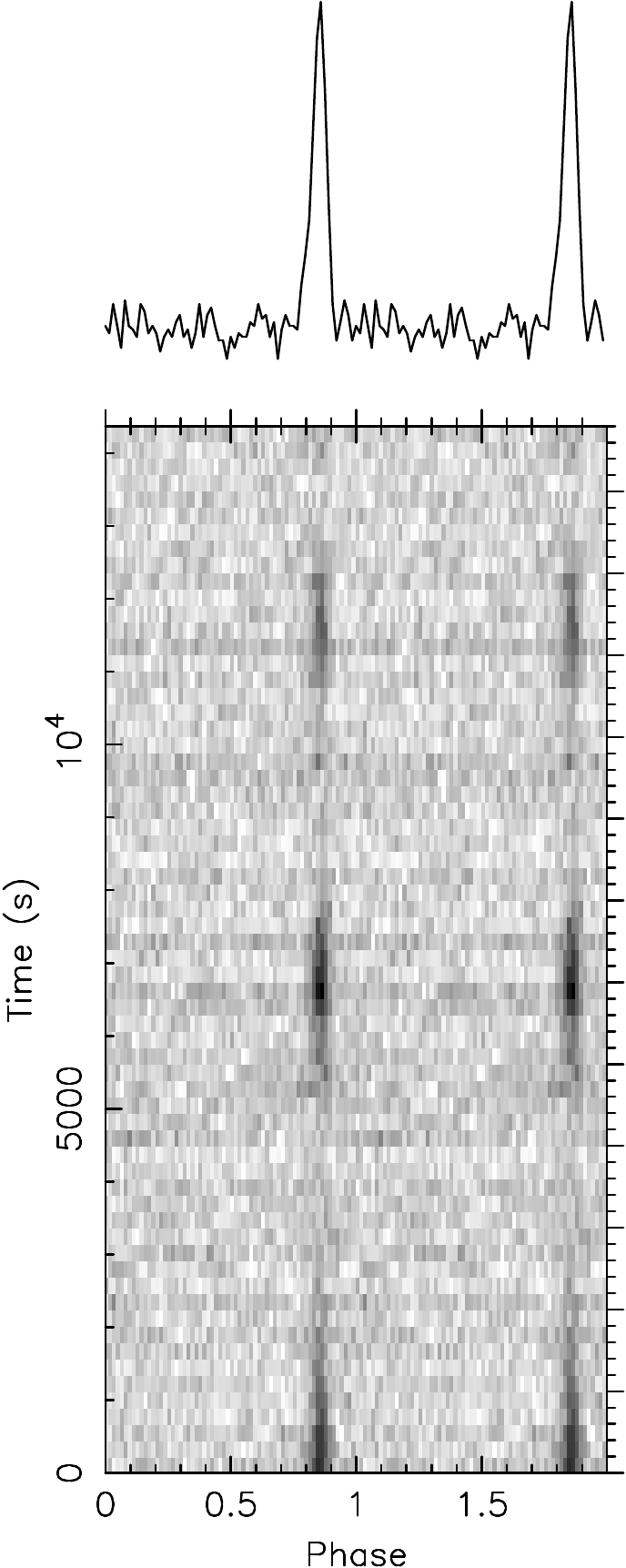}
  	\caption{Short-term flux density variations as seen in the isolated pulsar NGC~1851C in the UHF band (left panel) and in the binary pulsar NGC~1851E in the L band (right panel).}
  	\label{fig:scintillation_vs_time}
\end{figure}

\begin{figure}
\centering
	\includegraphics[width=0.45\columnwidth]{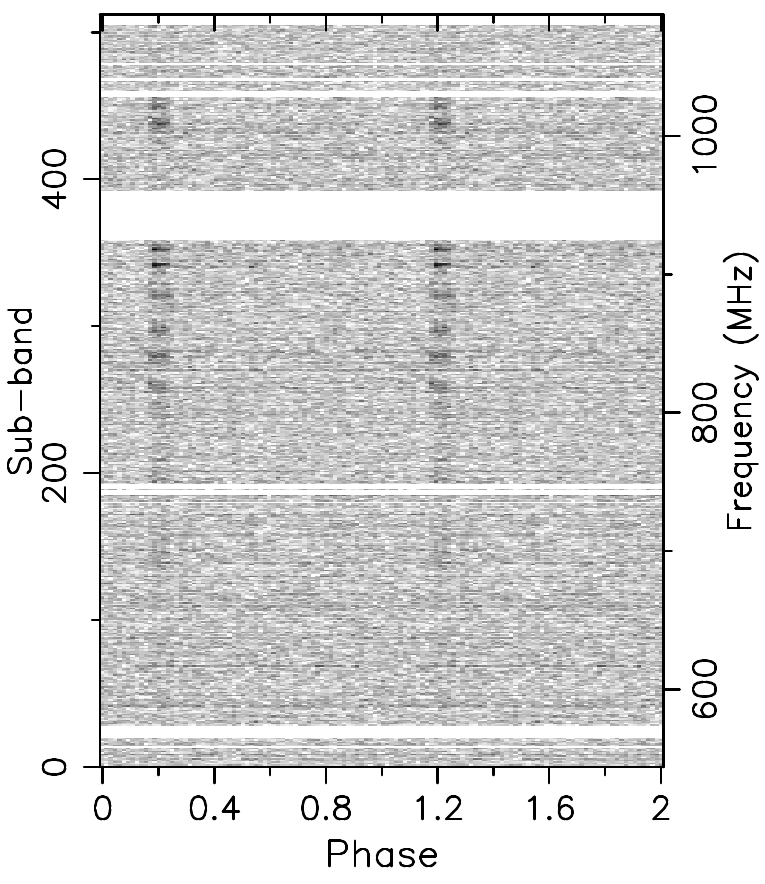}
	\qquad
	\includegraphics[width=0.45\columnwidth]{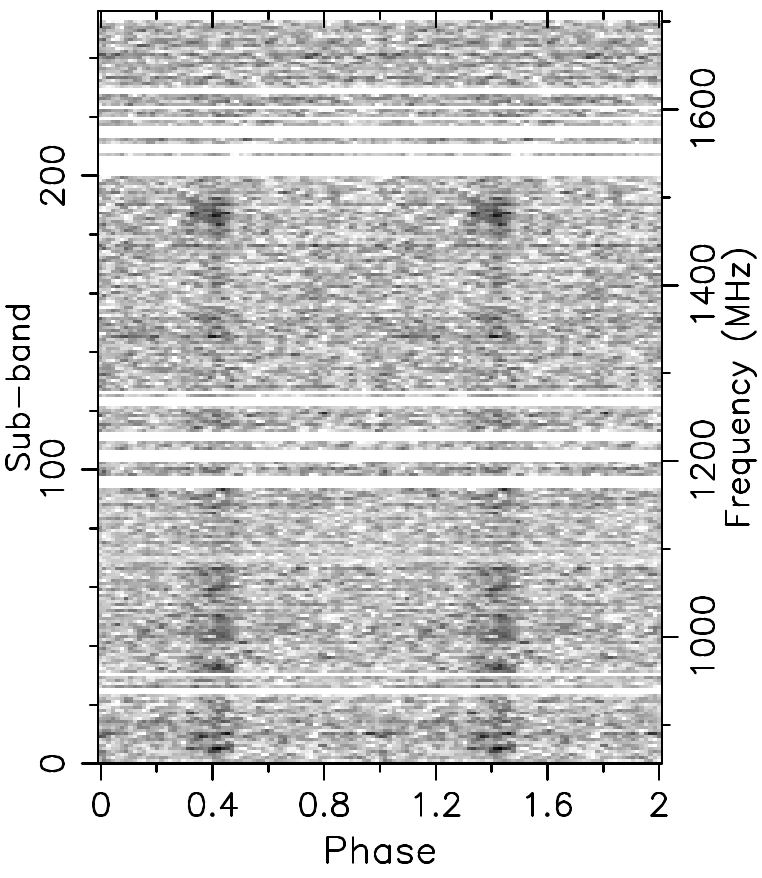}
  	\caption{Example of flux density variations as seen in the frequency domain in NGC~1851M in the UHF band over $\sim$30 minutes of integration time (left panel) and in NGC~1851B in the L band over $\sim$1 hour of integration time (right panel).}
  	\label{fig:scintillation_vs_freq}
\end{figure}

\section{Results}
\label{sec:results}

Our searches resulted in the discovery of 13 new pulsars in NGC~1851, which we designate from now on with the cluster's name and a letter, which reflects the order of discovery. No significant single pulses were found in any observation/beam, within the DM range explored (49.5$-$54.5~\dmunit).
All the discoveries are MSPs with spin periods in a very narrow range, 2.81 -- 6.63\,ms, the only exception being NGC~1851I, which is a 32.65-ms mildly recycled pulsar.  In Table \ref{tab:discoveries} we list the basic parameters derived for all the discoveries, while in Figure \ref{fig:integrated_profiles} we show their high-S/N, UHF-band integrated pulse profiles. In Table~\ref{tab:list_detections}, we list the detections of each of these pulsars in the MeerKAT observations. 

In the remainder of this section, we describe the characteristics of the new pulsars, and then outline some of their collective properties.

\subsection{Characteristics of the new pulsars}

About half of the new discoveries (NGC~1851B, C, J, K, M and N) are, based on our current understanding of their properties, isolated pulsars. This fraction is similar to that of Terzan 5, M28, NGC 6440 and NGC 6441. Pulsars C, J, K and M were also detected on a few occasions in archival GMRT data taken at 400 MHz, while both B and C were also seen a few times in the old GBT 350-MHz observations. However, all of these detections had low S/N.

One of these isolated pulsars, NGC~1851M, is extremely faint even in MeerKAT data: it was blindly  detected by the search pipeline in only four observations. Direct folding of all the MeerKAT observations yielded an additional four detections, all of which were extremely faint.
Because of this, we could not derive an improved position for this pulsar using \texttt{SeeKAT}. Nonetheless, we can assert that it must be located within $\sim$20-30 arcsec from the cluster center, as it is always detected with highest S/N in the central UHF beam of the TRAPUM tiling. We also note that NGC~1851M is the only pulsar, among the discoveries, that was not detected in the L band (see Table \ref{tab:list_detections}), suggesting that it has a particularly steep spectrum.

An additional five pulsars (NGC~1851F, G, H, I and L) are in binary systems with low-mass companions (as estimated from the mass function) and low-eccentricity orbits.
Pulsars F, G and H have orbital periods in the range $\sim 2 - 16.94$~d, which are typical of pulsar--white dwarf (WD) binaries.
Pulsars I and L, on the other hand, have orbital periods of $\sim$1~d, like those of a few ``redbacks''. These are a subclass of the ``spider'' binary pulsars (see \citealt{Roberts2013}), which are compact systems where the companion shows signs of being irradiated by the pulsar, and often exhibit other peculiar features. One of these is the presence of eclipses, where the radio pulses are either delayed or are undetectable for a good fraction of the orbit. Their absence in NGC~1851I and L, however, is a strong indication that the companion stars are not extended like the companions of redbacks, but instead are also low-mass WDs. Another characteristic of redbacks is the presence of long-term, erratic variations of the orbital period. We do not yet have enough data to decide on the stability of the orbit of any of these systems. Ultimately, optical observations of the companion stars are required to determine the actual nature of these systems.

At least two of these binaries, NGC~1851H and I, seem to have significant non-zero eccentricities. Despite being small, they are several orders of magnitude larger than similar pulsar--WD systems in the Galaxy. This is commonly observed in GCs and is due to perturbations caused by close stellar encounters, which are likely to happen in the crowded environment. Of this set of pulsars, G is the only pulsar detected in archival data, specifically in a few GMRT 400-MHz observations. 

The most interesting new pulsars are the two eccentric binary MSPs, NGC~1851D and E.
As for the other pulsars, no phase-connected timing solutions are yet available. However, their orbital parameters have been determined to high precision (see  Figure \ref{fig:NGC1851DE_Pobs_vs_MJD}).
These systems resemble NGC~1851A, the previously known binary pulsar in the cluster: they have large orbital eccentricities and massive companions which suggests they are very likely to be the result of secondary exchange encounters, in which the original low-mass He WD that recycled these pulsars was replaced by the current, more massive companion. NGC~1851D is in a 5.69-d orbit with an eccentricity of $e=0.86$. From its mass function, for a pulsar mass $\Mp=1.4$~\msun, we derive a companion mass  $\Mc>0.48$~\msun. Hence, we can preliminarily conclude that its companion is likely to be a massive Carbon-Oxygen WD. The pulsar has also been detected often in the existing GMRT and GBT data. Using these data we were able to extend its orbital solution to May 2005, when the GBT started observing NGC~1851A regularly \citep{Freire+2007}. 
NGC~1851E is in a 7.44-d, less eccentric ($e=0.71$)  orbit, but the companion is more massive: from the mass function, for $\Mp=1.4$~\msun, we obtain $\Mc>1.53$~\msun, which indicates that the companion is most likely a NS. Unlike D, it is not detectable in the bulk of the archival observations (which were made at lower frequencies, in order to time NGC~1851A, a steep spectrum pulsar), but it is detected in the few GBT observations made with the 820 MHz and S-band receivers \citep[as described in][]{Freire+2007}. 

All the newly found NGC~1851 pulsars were detected with the highest S/N in the central beam of the tiling of each TRAPUM observation. This is confirmed by the localizations obtained with \texttt{SeeKAT} as described in Section \ref{sec:localization}: all the maximum-likelihood pulsar positions are at $\lesssim 2$~arcsec from the nominal cluster center (Figure \ref{fig:localizations}). However, given the uncertainties, some objects could eventually display a larger offset. Much more accurate radio timing positions will soon be available once phase-connected solutions are achieved with future follow-up observations.

All the discoveries have DMs in the range $\sim$50.96\,$-$\,52.58~\dmunit. Including the previously known binary pulsar PSR~J0514$-$4002A, the DMs of the pulsars in NGC~1851 have an average of $51.82$~\dmunit\ and a standard deviation of $0.47$~\dmunit. The overall spread of $1.62$~\dmunit\ is just 3 per cent of the average value, in line with the DM distribution of all GCs known to host two or more pulsars.

For 10 out of the 14 pulsars in NGC~1851 we were able to measure their basic polarimetric properties as described in Section \ref{sec:polarimetry}. All the pulsars have best RM values between $-6$ and $-14$ \rmunit\ (see Table \ref{tab:discoveries}), although these are affected by large uncertainties due to the faintness of their polarized flux. While a few pulsars show high fractions of linear polarization ($>10$ per cent), pulsar K stands out for having an almost 100 per cent linearly polarized leading profile component (Figure \ref{fig:integrated_profiles}). We defer a more thorough analysis of the polarization properties of the NGC 1851 pulsars and their possible applications to a future publication.

\subsection{Flux density variations}
\label{sec:results_dms_fluxes}

In many clusters located well outside the Galactic plane (such as 47~Tuc and NGC~6752) the low average DM is often accompanied by large (more than 50 per cent in magnitude) and rapid (i.e. over time scales of < 1 h) variations in the apparent brightness of the pulsars, caused by diffractive scintillation \citep{Camilo+2000,Damico+2002}. This is also what we see in NGC 1851: all the new discoveries show very large flux density variations (sometimes by a factor of 3 or more) that can occur over time scales as short as a few tens of minutes (Figure \ref{fig:scintillation_vs_time}). When not averaged over long integration times, such variations manifest themselves as very small-bandwidth ``scintles'' in the frequency band (Figure \ref{fig:scintillation_vs_freq}), something that is particularly evident in the UHF-band observations. This is exactly what is expected in the case of diffractive scintillation, where the scintle size increases in both time and bandwidth in proportion to the fourth power of the frequency \citep{Rickett1977}.

What we see in the detections of the new pulsars can also be used to address a few hypotheses proposed by \citet{Freire+2007} to explain the large flux density variations that they saw in NGC~1851A. One of their argument was that those variations could be ``eclipses'' caused by gas outflowing from the companion star. This hypothesis can now be excluded, because these changes are seen in all our new discoveries, even those that are isolated pulsars.
The other possibility that they proposed was refractive scintillation caused by plasma structures in NGC~1851. However, refractive scintillation generally has long timescales of the order of weeks to years, not hours, and in addition, typically only modulates the measured flux densities by up to a factor of $\sim$2. It is for these reasons that \cite{Freire+2007} speculated on the presence of plasma ``lenses'' within NGC~1851 itself as a way of producing these changes on a much faster scale. However, in the case of refractive scintillation, the flux variations should be achromatic, i.e., independent of the radio frequency, which is not what we see in our MeerKAT data.

A detailed discussion of the scintillation phenomenon seen in the NGC~1851 pulsars is beyond the scope of this paper and will be addressed in a future study.

\begin{table}
\centering
\renewcommand{\arraystretch}{1.0}
\setlength{\tabcolsep}{0.15cm}
\footnotesize
\caption{Pulsars detected in each of the MeerKAT observations of NGC~1851. The star ($^\star$) indicates a detection possible only via direct folding, its absence indicates a detection using the searching pipeline.}
\label{tab:list_detections}
\begin{tabular}{cc}
\hline
Observation &  Pulsars detected  \\
\hline
01L      & A\phantom{$^{\star}$} B\phantom{$^{\star}$} C\phantom{$^{\star}$} D\phantom{$^{\star}$} E\phantom{$^{\star}$} F\phantom{$^{\star}$} G\phantom{$^{\star}$} H\phantom{$^{\star}$} I$^{\star}$\phantom{} \phantom{J$^{\star}$} K$^{\star}$\phantom{} L$^{\star}$\phantom{} \phantom{M$^{\star}$} N$^{\star}$\phantom{} \\

02L      & A\phantom{$^{\star}$} B\phantom{$^{\star}$} C\phantom{$^{\star}$} D\phantom{$^{\star}$} E\phantom{$^{\star}$} \phantom{F$^{\star}$} G\phantom{$^{\star}$} \phantom{H$^{\star}$} I\phantom{$^{\star}$} \phantom{J$^{\star}$} \phantom{K$^{\star}$} L$^{\star}$\phantom{} \phantom{M$^{\star}$} \phantom{N$^{\star}$}		\\

03L      & A\phantom{$^{\star}$} B\phantom{$^{\star}$} C\phantom{$^{\star}$} D\phantom{$^{\star}$} E\phantom{$^{\star}$} \phantom{F$^{\star}$} G\phantom{$^{\star}$} \phantom{H$^{\star}$} I\phantom{$^{\star}$} J$^{\star}$\phantom{} \phantom{K$^{\star}$} L$^{\star}$\phantom{} \phantom{M$^{\star}$} N$^{\star}$\phantom{}		\\

04U      & A\phantom{$^{\star}$} B\phantom{$^{\star}$} C\phantom{$^{\star}$} D\phantom{$^{\star}$} E\phantom{$^{\star}$} F$^{\star}$\phantom{} G\phantom{$^{\star}$} H\phantom{$^{\star}$} I\phantom{$^{\star}$} J\phantom{$^{\star}$} K\phantom{$^{\star}$} L$^{\star}$\phantom{} M$^{\star}$\phantom{} N\phantom{$^{\star}$}		\\

05U      & A\phantom{$^{\star}$} B\phantom{$^{\star}$} C\phantom{$^{\star}$} D\phantom{$^{\star}$} E\phantom{$^{\star}$} F\phantom{$^{\star}$} G\phantom{$^{\star}$} H$^{\star}$ I\phantom{$^{\star}$} J\phantom{$^{\star}$} K\phantom{$^{\star}$} L\phantom{$^{\star}$} M$^{\star}$\phantom{} N\phantom{$^{\star}$}		\\

06U      & A\phantom{$^{\star}$} B\phantom{$^{\star}$} C\phantom{$^{\star}$} D\phantom{$^{\star}$} E\phantom{$^{\star}$} F\phantom{$^{\star}$} G\phantom{$^{\star}$} H$^{\star}$\phantom{} I\phantom{$^{\star}$} J\phantom{$^{\star}$} K\phantom{$^{\star}$} L\phantom{$^{\star}$} M\phantom{$^{\star}$} N$^{\star}$\phantom{}		\\

07U      & A\phantom{$^{\star}$} B\phantom{$^{\star}$} C\phantom{$^{\star}$} D\phantom{$^{\star}$} E\phantom{$^{\star}$} F$^{\star}$\phantom{} G\phantom{$^{\star}$} \phantom{H$^{\star}$} I\phantom{$^{\star}$} J$^{\star}$\phantom{} K$^{\star}$\phantom{} L$^{\star}$\phantom{} \phantom{M$^{\star}$} N\phantom{$^{\star}$}		\\

08U    & A\phantom{$^{\star}$} B\phantom{$^{\star}$} C\phantom{$^{\star}$} D\phantom{$^{\star}$} E\phantom{$^{\star}$} F\phantom{$^{\star}$} G\phantom{$^{\star}$} H\phantom{$^{\star}$} I\phantom{$^{\star}$} J\phantom{$^{\star}$} K\phantom{$^{\star}$} \phantom{L$^{\star}$} M$^{\star}$\phantom{} N$^{\star}$\phantom{}		\\

09U    & A\phantom{$^{\star}$} B\phantom{$^{\star}$} C\phantom{$^{\star}$} D\phantom{$^{\star}$} E\phantom{$^{\star}$} \phantom{F$^{\star}$} G\phantom{$^{\star}$} H\phantom{$^{\star}$} I\phantom{$^{\star}$} J\phantom{$^{\star}$} K\phantom{$^{\star}$} L$^{\star}$ M$^{\star}$\phantom{} N\phantom{$^{\star}$}		\\

10U    & A\phantom{$^{\star}$} B\phantom{$^{\star}$} C\phantom{$^{\star}$} D\phantom{$^{\star}$} E\phantom{$^{\star}$} F\phantom{$^{\star}$} G\phantom{$^{\star}$} H\phantom{$^{\star}$} I$^{\star}$\phantom{} \phantom{J$^{\star}$} K$^{\star}$\phantom{} L\phantom{$^{\star}$} \phantom{M$^{\star}$} N\phantom{$^{\star}$}		\\

11U    & A\phantom{$^{\star}$} B\phantom{$^{\star}$} C\phantom{$^{\star}$} D\phantom{$^{\star}$} E\phantom{$^{\star}$} \phantom{F$^{\star}$} G\phantom{$^{\star}$} H\phantom{$^{\star}$} I\phantom{$^{\star}$} J$^{\star}$\phantom{} K\phantom{$^{\star}$} L\phantom{$^{\star}$} M\phantom{$^{\star}$} N$^{\star}$\phantom{}		\\

12U     &
A\phantom{$^{\star}$} B\phantom{$^{\star}$} C\phantom{$^{\star}$} D\phantom{$^{\star}$} E\phantom{$^{\star}$} F\phantom{$^{\star}$} G\phantom{$^{\star}$} H\phantom{$^{\star}$} I$^{\star}$\phantom{} \phantom{J$^{\star}$} K\phantom{$^{\star}$} L\phantom{$^{\star}$} M\phantom{$^{\star}$} N\phantom{$^{\star}$}		\\

13U    & A\phantom{$^{\star}$} B\phantom{$^{\star}$} C\phantom{$^{\star}$} D\phantom{$^{\star}$} E\phantom{$^{\star}$} \phantom{F$^{\star}$} G\phantom{$^{\star}$} H$^{\star}$\phantom{} I\phantom{$^{\star}$} J\phantom{$^{\star}$} K\phantom{$^{\star}$} L$^{\star}$\phantom{} M\phantom{$^{\star}$} N\phantom{$^{\star}$}		\\
\hline

\end{tabular}\\
\end{table}

%%%%%%%%%%%%%%%%%%%%%%%%%%%%%%%%%%%%%%%%%%%%%%%%%%%%%%%
%   5) DISCUSSION
%%%%%%%%%%%%%%%%%%%%%%%%%%%%%%%%%%%%%%%%%%%%%%%%%%%%%%%

\section{DISCUSSION}
\label{sec:discussion}

The 13 new discoveries presented here represent, thus far, the most successful search of a single GC with MeerKAT. They provide us with a far clearer view of the pulsar population of NGC~1851, and give us the opportunity to relate the pulsar characteristics with the cluster properties.

This population is unique in having three highly eccentric binary MSPs with massive companions, which are, in all likelihood, the result of secondary exchange encounters. Although several other clusters have a significant number of MSPs in eccentric binaries, like Terzan 5 and M28, those binaries have generally lower companion masses; this means that they are not as clearly identifiable as secondary exchange products.

As discussed by \cite{Verbunt_Hut1987}, the number of LMXBs and MSPs in a globular cluster depends primarily on the overall exchange encounter ratio, $\Gamma$, which depends on the cluster's structural parameters.
However, apart from their sizes, the pulsar populations in different clusters present strikingly different characteristics, like the fraction of binary pulsars, the presence of apparently young pulsars, and the numbers of systems formed in secondary exchange encounters.
As described by  \citet{Verbunt_Freire2014}, the so-called ``encounter rate for a single binary'', $\gamma$, explains these differences, but not others, like the distribution of MSP spin periods.

This parameter, which they normalize to the value of the cluster M4 ($\gamma_{\rm M4}$), is a measure of the probability for an already-formed binary system to undergo a close encounter with a third body, which could perturb the orbit, replace one of the stars, or even disrupt the whole system. In GCs with a large $\gamma$ --- especially those with collapsed cores --- we are much more likely to find slow, mildly recycled pulsars and products of secondary exchange interactions. This is because a NS that accretes matter in a LMXB phase has a smaller probability of being fully recycled and becoming an MSP before the system is disrupted due to a dynamical interaction, hence a significant fraction of pulsars in such clusters are partially recycled. Even if the system completes recycling, the resulting binary millisecond pulsar will also have a much higher probability of disruption. Because of such disruptions, the fraction of isolated pulsars should be larger for the clusters with larger $\gamma$.

This general picture is confirmed by the observations: in several core-collapsed, high-$\gamma$ clusters (NGC~6624, NGC~6752, NGC~6517, Terzan 1 and M15) the vast majority of the pulsars are isolated, and a good fraction are slow or mildly recycled; also, of the few known binaries in these clusters, a good fraction (PSR~B2127+11C in M15, \citealt{Jacoby+2006}, PSR J1807$-$2500B in NGC 6544, \citealt{Lynch+2012} and PSR J1823$-$3021G in NGC 6624,  \citealt{Ridolfi+2021}) are eccentric binary pulsars with massive companions that are clearly the result of secondary exchange encounters.

On the contrary, GCs with a smaller $\gamma$ provide an environment where a binary can survive, on average, for much longer. A LMXB will evolve undisturbed, and will almost inevitably evolve like the LMXBs in the Galaxy. The MSP population in such clusters will thus resemble that of the Galactic disk, with most of the pulsars being fully recycled MSPs, either isolated or in circular binaries with a He-WD, or with a more main sequence-like star, as in the case of the aforementioned ``spider'' systems. Striking examples of clusters of this type are 47~Tucanae \citep{Freire+2017} and M62 \citep{Lynch+2012}; these examples are even more striking because of their large total interaction rate, $\Gamma$, which is the reason why the absolute number of MSPs is so large.

Between these two groups are the clusters with intermediate values of $\gamma$. This is reflected in the observed pulsar populations, which have mixed characteristics. A major example is Terzan~5: of 39 pulsars that the latter is known to host, 19 are isolated and 20 are in binaries; almost all them are MSPs, but a few are mildly recycled; there are many circular binaries in pulsar-WD, black widow and redback systems, but also several wide, eccentric, and sometimes massive, binaries. Almost identical considerations apply to M28, as well as NGC~6440 and NGC 6441.

With an encounter rate per binary of $\gamma = 12.44\,\gamma_{\rm M4}$, NGC~1851 is also an intermediate-$\gamma$ GC. This is reflected in the characteristics of its pulsar population: a) The number of isolated pulsars (6) is comparable to the number of binary pulsars (8); b) all the pulsars are MSPs, with the exception of NGC~1851I, which is a mildly-recycled pulsar; c) among the binaries, three (pulsars A, D and E) are extremely eccentric and clearly result from secondary exchange encounters, while all the others look to be ``standard'' pulsar--WD systems.

Notwithstanding, NGC~1851 also shows some marked differences with some of the other intermediate-$\gamma$ clusters. Firstly, similarly to NGC 6441 but at odds with Terzan 5, NGC~6440 and M28, none of the eight binary pulsars are definite spiders (although I and L could be wide-orbit, non-eclipsing redbacks). However, this could be due to the small number statistics and to the fact that spider pulsars are notoriously elusive: the presence of long eclipses and their extremely compact orbits make them challenging targets for search algorithms. Notwithstanding, this is unlikely to represent the full explanation: M28 has the same number of known pulsars (14), but fully half of its pulsar population consists of ``spiders'': five black widows (M28G, J, L, M, N) and two redbacks, M28H and I, the latter being a transitional MSP system \citep{Papitto+2013}. More observations, especially with the S-band receivers of MeerKAT, and the use of more sophisticated search techniques may eventually reveal some spiders in NGC~1851.  

Another interesting aspect is that all the 14 pulsars in NGC~1851 are within or just outside the core. This situation is similar to that of M28, where, with the exception of M28F, all pulsars are also very close to the center \citep{Begin2006}. Although some concentration is to be expected from mass segregation, which makes heavier objects (such as NSs) sink towards the center of the cluster, it is certainly much more centrally condensed than the pulsar populations of Terzan 5 and 47 Tucanae, which are generally understood to be in dynamical equilibrium with the remaining stellar populations in the cluster (see e.g., \citealt{Grindlay+02} and references therein). The reasons for this peculiarity of NGC~1851 and M28 are not clear: either their core radii have not been accurately measured, or the pulsars are confined by a deeper, narrower gravitational potential caused by a central concentration of massive objects that do not affect the overall distribution of luminous stars. A detailed investigation of this issue will be carried out once timing solutions are available for all pulsars in NGC~1851.

\section{Conclusions and prospects}
\label{sec:conclusions}

In this paper, we have presented 13 new pulsars found by TRAPUM in the GC NGC~1851 using the MeerKAT radio telescope. Six of these are solitary millisecond pulsars, while the other seven are part of binary systems. Among these discoveries, there are a few noteworthy pulsars. NGC~1851D is a MSP in a wide orbit with a heavy WD companion. The system has an eccentricity of $e=0.86$, which makes it the 5th most eccentric binary known. NGC~1851E is also a MSP in a very eccentric ($e=0.71$) binary with a NS as companion.
In all likelihood, both binaries formed in secondary exchange encounters where the original star that recycled the pulsar was replaced by the current, heavier companion. NGC~1851 is a cluster with intermediate interaction rate per single binary, $\gamma$. This is reflected in its pulsar population, which shows properties that are in between those found in core-collapsed and dynamically relaxed clusters.

The large ensemble of pulsars in NGC~1851 opens up interesting scientific possibilities for future studies.
One of the obvious extensions of the work presented here using the current TRAPUM data, will be the search for additional pulsars in all the beams beyond the half-mass radius, which have so far been excluded due to computational and time constraints. This could reveal, for instance, systems that may have been ejected from the core during secondary exchange interactions, and that have not had time yet to sink back in the core (these types of objects are more common in GCs with a large $\gamma$, see \citealt{Verbunt_Freire2014}). To complement this, all the beams, starting from the inner ones, will also be searched using more sophisticated (but often more computationally demanding) methods, such as jerk searches \citep{Andersen_Ransom2018}, phase-modulation searches \citep{Ransom+2003}, template-banking algorithms over 3 or 5 Keplerian parameters \citep{Balakrishnan+2022}, as well as fast-folding algorithms \citep{Morello+2020}; the first three have a higher potential of unveiling fast pulsars in extremely compact binaries, while the latter is more sensitive to very slow, isolated pulsars. With such searches, we will know whether the intriguing lack of ``spiders'' is real, or a result of current limitations in our search techniques. Given the intermediate structural characteristics of NGC~1851, the likelihood of finding both types of pulsars is not negligible. 

The polarization properties of the pulsars presented in this work will be further studied as more follow-up observations are taken in full-Stokes mode. If  more precise RM measurements are possible, these could potentially be exploited to probe the Galactic magnetic field at arcsecond scales in a different direction than that explored by \cite{Abbate+2020b}, who used the pulsars in 47 Tucanae.

With a relatively small number of additional observations, we can expect to derive phase-connected timing solutions for all the newly found pulsars. These will immediately give access to a much more precise determination of their projected positions, which will allow us to look for possible X-ray counterparts in \emph{Chandra} data, as well for the optical counterparts of the binary pulsars in archival \emph{HST} and future \emph{JWST} data.

Continued timing in the coming years will enable the measurement of the proper motions and of the higher-order spin period derivatives for all pulsars. This will allow us to constrain the cluster gravitational potential and probe the presence of non-luminous matter (such as a possible intermediate-mass back hole) at its center \citep[e.g.][]{Freire+2017,Abbate+2018,Abbate+2019}. This and a more in-depth study of the scintillation patterns observed in the pulsars could shed light on the possible presence of an intracluster medium, as already seen in 47 Tuc \citep{Freire+2001b,Abbate+2018}.

For the binary pulsars (especially for the very eccentric systems, NGC 1851D and E) the long-term timing will yield precise measurements of the rate of advance of periastron ($\dot{\omega}$). Additional post-Keplerian parameters are desirable, particularly for determining the masses of the individual components. A detection of the Shapiro delay, even at a very low level, can, when combined with a measurement of $\dot{\omega}$, yield precise masses \citep[e.g.~NGC~6544B in][]{Lynch+2012}; however, this is especially difficult in this case given how faint the pulsars are, and how low is their timing precision. For NGC~1851E, it will be possible to measure the Einstein delay ($\gamma_{\rm E}$), but, as in the case of NGC~1851A, this will take at least a decade \citep{Ridolfi+2019}. This is not likely to happen for NGC~1851D: because of its longitude of periastron ($\sim 358 \deg$, close to $0 \deg$), other geometric effects caused by the proper motion of the system will make $\gamma_{\rm E}$ undetectable (see detailed discussion in subsection 3.7 of \citealt{Ridolfi+2019}). 
For all the binaries, the measurements of the orbital period derivative would give access to reliable estimation of the intrinsic spin-down rate, surface magnetic filed and characteristic age of the pulsars.

Many of these goals will greatly benefit from archival data taken at other radio facilities (particularly GBT and GMRT) if the pulsars are also frequently detected in them once their phase-connected solutions are determined.

The discoveries presented in this paper are among the first made by the TRAPUM GC pulsar survey. The latter exploits the outstanding sensitivity of all the 64 antennas available at MeerKAT, as well as advanced beamforming techniques to search for pulsars in a large portion of 28 selected GCs. As more and more clusters are observed, we expect to significantly increase the known population of pulsars in several other GCs, allowing new studies and a much deeper understanding of the conditions and the physical processes that drive the evolution of these stellar systems.

\begin{acknowledgements}
%%%%%%%%%%%%%%%%%%%%%%
% TELESCOPES
%%%%%%%%%%%%%%%%%%%%%%
% MeerKAT
The MeerKAT telescope is operated by the South African Radio Astronomy Observatory, which is a facility of the National Research Foundation, an agency of the Department of Science and Innovation. SARAO acknowledges the ongoing advice and calibration of GPS systems by the National Metrology Institute of South Africa (NMISA) and the time space reference systems department of the Paris Observatory. MeerTime data is housed on the OzSTAR supercomputer at Swinburne University of Technology. The OzSTAR program receives funding in part from the Astronomy National Collaborative Research Infrastructure Strategy (NCRIS) allocation provided by the Australian Government. 
PTUSE was developed with support from  the Australian SKA Office and Swinburne University of Technology.
%% GBT / NRAO
The National Radio Astronomy Observatory is a facility of the National Science Foundation operated under cooperative agreement by Associated Universities, Inc. 
The Green Bank Observatory is a facility of the National Science Foundation operated under cooperative agreement by Associated Universities, Inc.
%% GMRT
We thank the GMRT staff for help with the observations that were used in this work. The GMRT is operated by the National Centre for Radio Astrophysics (NCRA) of the Tata Institute of Fundamental Research (TIFR), India.
%%%%%%%%%%%%%%%%%%%%%%%
%% COMPUTING
%%%%%%%%%%%%%%%%%%%%%%%
This work was partly performed on the OzSTAR national facility at Swinburne University of Technology. The OzSTAR program receives funding in part from the Astronomy National Collaborative Research Infrastructure Strategy (NCRIS) allocation provided by the Australian Government.
The authors also acknowledge MPIfR funding to contribute to MeerTime infrastructure. TRAPUM observations used the FBFUSE and APSUSE computing clusters for data acquisition, storage and analysis. These clusters were funded and installed by the Max-Planck-Institut für Radioastronomie and the Max-Planck-Gesellschaft.
%%%%%%%%%%%%%%%%%%%%%%
% AUTHORS
%%%%%%%%%%%%%%%%%%%%%%
%INAF
APo, AR and MBu gratefully acknowledge financial support by the research grant ``iPeska'' (P.I. Andrea Possenti) funded under the INAF national call Prin-SKA/CTA approved with the Presidential Decree 70/2016. APo, AR, MBu also acknowledge support from the Ministero degli Affari Esteri e della Cooperazione Internazionale - Direzione Generale per la Promozione del Sistema Paese - Progetto di Grande Rilevanza ZA18GR02. SMR is a CIFAR Fellow and is supported by NSFP hysics Frontiers Center awards 1430284 and 2020265.
%%Scott's acks
SMR is a CIFAR Fellow and is supported by the NSF Physics Frontiers Center award 1430284.
%%Federico's acks
FA, MK, and PVP gratefully acknowledges support from ERC Synergy Grant ``BlackHoleCam'' Grant Agreement Number 610058.
%% MPIfR acks
AR, TG, PCCF, VVK, MK, FA, EDB, PVP, DJC, WC and APa 
acknowledge continuing valuable support from the Max-Planck Society.
%Ben
BWS and MCB acknowledge funding from the European Research Council (ERC) under the European Union’s Horizon 2020 research and innovation programme (grant agreement No. 694745).
%% Laila
LV acknowledges financial support from the Dean’s Doctoral Scholar Award from the University of Manchester.
%%Rene
RPB acknowledges support from the ERC
under the European Union’s Horizon 2020 research and innovation programme (grant agreement no. 715051; Spiders).
%Matthew
MBa acknowledges support through ARC grant CE170100004.
%Referee
We are grateful to the anonymous referee for useful comments, which helped improve the manuscript.
\end{acknowledgements}

\bibliographystyle{aa} % style aa.bst
\bibliography{NGC1851_discoveries} % your references Yourfile.bib
\end{document}